%% file: main.tex
\documentclass[aps,prd,reprint,superscriptaddress,floatfix]{revtex4-2}
\input{preamble}
\usepackage{graphicx}
\usepackage{rotating}
\usepackage{multirow}
\begin{document}
\title{Towards a Global CE$\nu$NS Picture}

\input{authors}
\date{\today}

\input{abstract}

\maketitle

\section{Introduction}
\label{sec:introduction}
\input{sections/intro}

\section{The $\cevns$ Cross-section}
\label{sec:cross_section}
\input{sections/theory}

\section{The Predictive Framework}
\label{sec:predictive_model}
\input{sections/framework_intro}
\subsection{The Cross-Section}
\label{sec:xsec}
\input{sections/framework}

\subsection{Neutrino Flux}
\label{sec:neutrino_flux}
\input{figures/flux_envelope_fig}
\input{sections/neutrino_flux}
\subsection{Detector Response}
\label{sec:detector_response}
\input{sections/response.tex}
\subsection{The Forward Model}
\label{sec:fwd_model}
\input{sections/fwd_model.tex}
\input{figures/corner_copy.tex}
\input{figures/cevns_xsec_fig.tex}

\begin{turnpage}
\begin{table*}
    \input{tables/params}

\input{tables/norm_fit_data}

\input{tables/param_fit_data}
\end{table*}
\end{turnpage}

\section{Results}
\label{sec:results}
\input{sections/results.tex}

\subsection{Example Application}\label{sec:lz_test}
\input{sections/LZ_test}

\section{Summary}
\label{sec:summary}
\input{sections/summary}
\section*{Acknowledgments}

We would like to thank Oliver Schulz for helpful discussion and help regarding the inference part.
This work was funded by the Deutsche Forschungsgemeinschaft (DFG, German Research Foundation) under 
Germany’s Excellence Strategy -- EXC-2094/2 -- 390783311.

\appendix
\label{sec:appendix}
\section{COHERENT l-Ar [CENNS10] Implementation}
\label{app:lAr}
\input{sections/COHERENT_Ar.tex}

\section{COHERENT CsI[Na] Implementation}
\label{app:CsI}
\input{sections/COHERENT_CsI.tex}

\bibliography{refs}

\end{document}

%% file: preamble.tex
\usepackage{graphicx,amsmath,mathtools}
\usepackage{epsfig,subfigure}
\usepackage{array, dcolumn}
\usepackage{amsmath, amssymb}
\usepackage{hyperref}
\usepackage[normalem]{ulem} 
\usepackage{xcolor}

\def\beq{\begin{equation}}
\def\eeq{\end{equation}}
\def\bea{\begin{eqnarray}}
\def\eea{\end{eqnarray}}

\def\3Eqs#1#2#3{Eqs.\ (\ref{#1}), (\ref{#2}) and (\ref{#3})}

\def\within#1to#2/{#1\mbox{ to }#2}
\def\lbullet#1,#2,#3,#4/{\Text(#1,#2)[c]{$\bullet$}\Line(#1,#2)(#3,#4)}
\def\textcite#1{Ref.~\cite{#1}}

\def\hd0ino{\tilde{h^0_{d}}}

\def\hu0ino{\tilde{h^0_{u}}}

\def\e2spl{\ensuremath{^\clubsuit}}

\def\nn{\nonumber\\*}

\newcommand{\cevns}{\text{CE$\nu$NS}}



%% file: authors.tex
\author{Heerak Banerjee}
\email{banerjee@mpp.mpg.de}
\affiliation{Max Planck Institute for Physics, Boltzmannstr. 8, D-85748 Garching, Germany}
\affiliation{Excellence Cluster ORIGINS, Boltzmannstr. 2, D-85748 Garching, Germany}
\author{Philipp Eller}
\email{philipp.eller@tum.de}
\affiliation{Technical University Munich, James-Franck-Str. 1, D-85748 Garching, Germany}
\affiliation{Excellence Cluster ORIGINS, Boltzmannstr. 2, D-85748 Garching, Germany}

%% file: abstract.tex
    \begin{abstract}
We present a general framework for extracting and interpreting the Coherent Elastic Neutrino-Nucleus Scattering ($\cevns$) differential cross-section from experimental data. A generalized parametric basis, constructed from common kinematic terms recurring in generalized neutrino interaction models, is employed to numerically constrain the shape and normalization of the observed $\cevns$ response. The framework was applied to the COHERENT CsI and liquid argon datasets using Bayesian inference, yielding normalization constraints comparable to those reported by the Collaboration while providing additional constraints on the correlated spectral shape. A combined analysis was also performed, exploiting the complementary sensitivities of the two datasets to constrain features of the differential cross-section that are less accessible to either dataset individually. The framework has been implemented in the open-source Julia package \texttt{Newtrinos.jl}, which supports modular forward modeling, parallel computation, and automatic differentiation.
\end{abstract}

%% file: sections/intro.tex
Coherent Elastic Neutrino-Nucleus Scattering ($\cevns$) remained experimentally unexplored for over four decades after its prediction in 1974\,\cite{Freedman:1973yd}. The process is characterized by the coherent enhancement of the cross-section arising from the in-phase scattering of an incident neutrino off all nucleons in a target nucleus, resulting in an approximate scaling with the square of the number of participating nucleons. $\cevns$ is predominantly mediated by the neutral-current Weak interaction, with typical cross-sections in the femtobarn range, despite this enhancement. Furthermore, its low-energy kinematics lead to a rapid suppression of the cross-section as the transferred momentum approaches the inverse nuclear radius. These features make its experimental observation challenging. The first observation by the COHERENT Collaboration on CsI[Na]\cite{Akimov:2017ade} therefore provided an important validation of both the SM prediction and the underlying interplay of particle, hadronic and nuclear physics.

The $\cevns$ response depends on physical inputs at three distinct length scales, with contributions at each scale depending on those at shorter scales. At the shortest scale, the neutrino-quark couplings depend on the Weak coupling strength and mixing angle, as well as potential beyond-the-Standard-Model (BSM) contributions. At the nucleonic scale, the response is determined by hadronic matrix elements whose structure depends on the nature of the $\nu-q$ current. Finally, the nuclear response depends on the target nucleus and the contributions at the preceding scales. These inputs are generally not factorizable in the observable response, except within restricted kinematic domains or under simplifying assumptions. Variations in any of them can affect both the shape and normalization of the expected $\cevns$ spectrum. Disentangling their contributions, therefore, requires a combined analysis of measurements from complementary targets, neutrino sources, and detector configurations, with sufficient statistical precision to resolve spectral features.

This has been realized through several measurements of $\cevns$ following its first observation. The COHERENT Collaboration has reported additional measurements using CsI~\cite{COHERENT:2021xmm}, liquid argon (l-Ar)~\cite{COHERENT:2020iec} and Germanium~\cite{COHERENT:2026yje}, all with a stopped-pion neutrino source. The Dresden-II reactor experiment has also reported preliminary observations of reactor antineutrinos using Germanium~\cite{Colaresi:2021kus}. Other experiments have pursued measurements using Silicon, Germanium, Calcium Tungstate, Sodium Iodide and other targets, with different neutrino sources and detection technologies~\cite{CCM:2021leg,CONNIE:2024pwt,CONUSCollaboration:2024kvo,CONUS:2024lnu,Mondal:2026vvx,TEXONO:2024vfk,Belov:2015ufh,Ackermann:2025obx,NUCLEUS:2026pnv,Bailly-Salins:2026inm}. These efforts have been accompanied by increments in neutrino flux densities and detector exposure alongside better characterization of detector response, backgrounds and source profiles.

Theoretical analyses have similarly sought to constrain the underlying physical inputs using available measurements~\cite{AtzoriCorona:2025xwr,AtzoriCorona:2024vhj,Cadeddu:2023tkp,AtzoriCorona:2023ktl,Cadeddu:2017etk,Cadeddu:2018dux,Papoulias:2019lfi,Huang:2019ene,Coloma:2020nhf,Cadeddu:2021ijh,Miranda:2020tif,Cadeddu:2020lky}. These include low-energy tests of the SM neutrino sector, particularly the Weak interaction, which remains relatively poorly tested at the momentum transfers relevant to $\cevns$ ($Q^2 \lesssim 100$ MeV). Potential BSM contributions may also modify the effective SM parameters at these scales. Measurements of $\cevns$ can therefore constrain the Weak mixing angle ($\sin^2\theta_W$), neutrino-quark couplings and the neutrino charge radius. The nuclear response introduces an additional source of uncertainty in these measurements, but also allows the same data to constrain nuclear parameters. In particular, the Weak nature of the process makes it sensitive to the neutron distribution within the target nucleus, a topic of considerable theoretical and experimental study.

At the particle-interaction scale, $\cevns$ is also sensitive to several BSM scenarios that can modify the shape and normalization of the expected spectrum. These include non-standard neutrino interactions, neutrino electromagnetic properties, sterile neutrinos, and hypothetical light mediators. The response may also depend on the nature of the neutrino, specifically whether it is a Dirac or Majorana particle. Their effects on $\cevns$ and the constraints obtainable from existing measurements have been studied extensively~\cite{Drukier:1983gj,Giunti:2015nel,Cadeddu:2020nbr,Shoemaker:2017lzs,Liao:2017uzy,Abdullah:2018ykz,AristizabalSierra:2018eqm,Chatterjee:2024vkd,Barranco:2005yy,Scholberg:2005qs,Barranco:2007tz,Dutta:2015vwa,Lindner:2016wff,Dent:2016wcr,Coloma:2017egw,Anderson:2012pn,Dutta:2015nlo,Kosmas:2017zbh,Dodd:1991ni,Kosmas:2015sqa,deNiverville:2015mwa,Coloma:2020gfv,Coloma:2020nhf,Dutta:2020vop,Dutta:2020enk,Miranda:2020zji,Dutta:2019eml,Miranda:2019wdy,Shoemaker:2021hvm,Denton:2018xmq,Banerjee:2021laz,Kelly:2026avh,Binh:2026ehh,Bisset:2026fqy,AristizabalSierra:2026rlo,Carey:2025byg,Behera:2025hdv,Breso-Pla:2025pds,TEXONO:2025sub,Liao:2024qoe,Li:2024gbw,Chatterjee:2024vkd,DeRomeri:2023cjt,Breso-Pla:2023tnz,Calabrese:2022mnp}.

With the existence of $\cevns$ experimentally established, the next challenge is to characterize its observed spectrum. This requires a forward model capable of accommodating deviations in both the normalization and shape of the SM prediction. To separate particle physics from uncertainties in the nuclear response and neutrino flux, the extraction should not depend strongly on specific assumptions about these inputs. The resulting spectrum should also permit interpretation within different nuclear models and BSM scenarios. This is particularly relevant to dark matter direct-detection experiments, where $\cevns$ constitutes an irreducible background from cosmic and solar neutrinos, commonly referred to as the ``neutrino fog''~\cite{Akerib:2022ort,OHare:2016pjy,Monroe:2007xp}. At sufficiently high sensitivities, such events become indistinguishable from nuclear recoils induced by hypothetical dark matter particles. An experimentally extracted $\cevns$ spectrum, independent of specific assumptions about the neutrino flux and underlying interaction, would provide a basis for estimating this background across different experimental configurations. The framework must therefore also accommodate different neutrino flux profiles and allow the combination of datasets from multiple targets, sources, and detector technologies.

This work presents a general framework for extracting the $\cevns$ differential cross-section numerically from experimental data and interpreting the resulting constraints in terms of the underlying particle and nuclear physics. General descriptions of the differential cross-section in terms of neutrino interactions~\cite{Lindner:2016wff} and effective field theories (EFTs)~\cite{Hoferichter:2020osn} are already available. Ref.~\cite{Hoferichter:2020osn} also discusses the nuclear response associated with the different EFT operators and provides explicit parameterizations of the corresponding response functions. However, the extraction of an observed spectrum primarily depends on the functional form of the forward model. The differential cross-sections associated with different operators in $\cevns$ EFTs can be expressed as combinations of a relatively small number of common terms, polynomial in the primary kinematic variables: the incident neutrino energy ($E_\nu$), nuclear recoil energy ($E_R$), and target nuclear mass ($M_N$). We investigate the use of these terms as a basis for a general parameterization of the differential cross-section, with independent coefficients controlling its shape and normalization. The resulting model can accommodate deviations from the fixed SM spectral shape while retaining a direct connection to a variety of theoretical descriptions. Depending on the choice of parameter priors, the resulting posterior distributions can be interpreted either as model-independent constraints or as constraints on specific physical models. The $\cevns$ response includes a dominant common shape from the shared particle physics across nuclei as well as a target-dependent component. The framework makes this distinction explicit through its interpretation.

We apply the framework to the publicly available COHERENT datasets on CsI and liquid argon~\cite{Akimov:2020czh,COHERENT:2021xmm}. Bayesian fits with uninformed flat priors were found to constrain the normalization of the extracted spectrum to a level comparable to that reported by the Collaboration, while also providing additional insights into correlated variations in spectral shape. The analysis yields a model-agnostic, numerical extraction of the $\cevns$ differential cross-section as a function of both recoil and incident neutrino energies for the target nuclei considered. A combined fit to both datasets was found to constrain features of the spectrum that are less accessible to either dataset individually, exploiting their complementary sensitivities. Finally, the extracted cross-section was used in an example application to examine the recently reported event at the LUX-ZEPELIN (LZ) experiment~\cite{LZ:2026axp} in terms of a possible stray $\cevns$ event. The available $\cevns$ cross-section within the corresponding nuclear-recoil window was small, as expected, but non-zero and slightly larger than the SM prediction.

The framework is implemented in the open-source Julia package \texttt{Newtrinos.jl}. Its modular architecture separates the physical differential cross-section, neutrino flux and detector response into independently constructed generalized arrays. The implementation supports parallel computation and automatic differentiation, providing a basis for integration into larger-scale analysis and optimization workflows.

The remainder of this paper is organized as follows. Sec.~\ref{sec:cross_section} discusses the $\cevns$ cross-section within and beyond the SM, while Sec.~\ref{sec:xsec} describes its implementation in the framework. Sec.~\ref{sec:neutrino_flux} discusses the relevant neutrino flux profiles and their incorporation into the predictive model. Sec.~\ref{sec:detector_response} describes the construction of detector response matrices and their interaction with the other components. Sec.~\ref{sec:fwd_model} presents the complete forward model. Sec.~\ref{sec:results} discusses the results from the individual and combined dataset analyses, as well as the application to the LZ event. Finally, Sec.~\ref{sec:summary} summarizes the findings and discusses future directions.

%% file: sections/theory.tex
$\cevns$ is inherently a low-energy, neutral-current, nuclear process. The condition of elasticity requires that the involved nucleus remain unchanged and in its ground state throughout the interaction. The entire transferred momentum is presumed to be observable as a nuclear recoil kinetic energy. 
The crucial distinction of "coherence" in $\cevns$ generally implies that the interaction may proceed in tandem with {\it all} the nucleons, such that the scattering amplitudes off individual nucleons sum {\it in-phase},
\beq
\mathcal{M} = \sum_{k=1}^A \mathcal{M}_k e^{i\Vec{q}\cdot\Vec{r_k}} \approx \sum_{k=1}^A \mathcal{M}_k.
\eeq
The interaction is said to be approximately coherent as long as this condition may be satisfied, or, equivalently,
\beq
q R_N \ll 1 \label{eqn:coherence_cond},
\eeq
where $q$ is the momentum transferred by a neutrino incident on a nucleus of radius $R_N$. For a nucleus of mass $M_N$ recoiling with energy $E_R$, the transferred momentum is approximately $q^2 \approx 2 M_N E_R$ as long as $E_R/M_N \ll 1$. We have, in addition, a purely kinematic bound relating the nuclear recoil energy $E_R$ that may be imparted to a nucleus of mass $M_N$ by a neutrino of energy $E_\nu$,
\begin{equation}
    \frac{M_N E_R}{2E_\nu^2}\leq 1. \label{eqn:kinematic_cond}
\end{equation}
The former implies a soft upper bound on nuclear recoil energy, beyond which the process may not remain completely coherent. The latter puts a lower bound on the neutrino energies, below which, depending upon the experimental threshold, the process may not be observable.

The collective participation of all the nucleons within the coherent regime leads to an enhancement of the scattering cross-section by a factor of $\sim \mathcal{N}_P^2$, where $\mathcal{N}_P$ is the number of participating nucleons. This is the origin of the ``coherent enhancement" in $\cevns$ and other coherent processes. The coherence of the interaction depends, in addition, on the nature of the underlying quark-level current and the nuclear response it elicits. In the SM, for example, the Vector current does elicit a coherent response, while the corresponding Axial current, which couples only to the residual nuclear magnetic moment, does not. The condition of Eqn.\ref{eqn:coherence_cond} is therefore a necessary but not sufficient condition for $\cevns$ to occur.  The degree of coherence, or the lack thereof, is captured ultimately by the nuclear form factors $F_X(q^2)$, which are functions of the transferred momentum $q^2$, the nuclear structure and the strengths of the respective currents. 

A naive picture of the general process may thus be formed: For $\cevns$ to be observable, one requires a sufficiently low-energy neutrino flux that leads to an appreciable coherent enhancement, yet energetic enough to be detected by the detectors' threshold energy. A larger nucleus typically leads to a greater coherent enhancement, but also an earlier onset of decoherence, not to mention smaller recoil energies relative to their more compact counterparts. Observations within the decoherence regime may provide a better handle on the nuclear response, while those in the coherent regime, where the form factors are effectively unity, may better resolve the microphysics. The optimal choice of a target nucleus and neutrino source is therefore a delicate balance among these competing factors, in addition to the ubiquitous technical considerations. Comparison of observations across multiple targets and sources, alongside inputs from other experiments and theoretical calculations, may also prove crucial to forming a more complete picture of the underlying physics. The categorical estimation of this physics could be indispensable in other domains where, for example, $\cevns$ may form a primary background component, or where it may be limited by nuclear response systematics. The following sections will describe the $\cevns$ process in more detail, first within the SM, then in a broader BSM paradigm. Finally, we describe a framework that may make optimality evaluations easier while also facilitating the global reinterpretation and analysis of $\cevns$ data in a relatively model-agnostic manner.

\subsection*{$\cevns$ in the Standard Model}
In the SM, the overwhelmingly dominant $\cevns$ channel proceeds exclusively via the exchange of a $Z$-Boson between the incoming neutrino and the target nucleus. The interaction is insensitive to neutrino flavor. The process entails the scattering of an incoming neutrino with energy $E_\nu$ off a nucleus of mass $M_N$ and ground-state angular momentum $J_N$, inducing, solely, a nuclear recoil energy $E_R$.
Up to the first sub-leading order in the SM, the $\cevns$ differential cross-section can be written down as~\cite{Freedman:1973yd, Drukier:1983gj, Lindner:2016wff, AristizabalSierra:2018eqm, Hoferichter:2020osn, Chatterjee:2024vkd},
\begin{eqnarray}
\frac{d\sigma}{dE_R}(q^2)
  = \frac{G_F^2}{4\pi}
    M_N
    \times \left(\left\vert F_W(q^2)\right\vert^2\mathcal{K}_V + F_A(q^2)\mathcal{K}_A\right), \label{eqn:sm-rate}
\end{eqnarray}
where,
\begin{eqnarray}
    \mathcal{K}_V &=& 1 - \frac{M_N E_R}{2E_\nu^2} - \frac{E_R}{E_\nu} + \frac{E_R^2}{2E_\nu^2},\nn
    \mathcal{K}_A &=& 1 + \frac{M_N E_R}{2E_\nu^2} - \frac{E_R}{E_\nu} + \frac{E_R^2}{2E_\nu^2},\nn
    F_W(q^2=0) &=& \mathcal{Z} Q^p_W + \mathcal{N} Q^n_W,\nn
    F_A(q^2=0) &=& \frac43 \frac{J_N+1}{J_N}\left(\Delta^p_u - \Delta^p_d\right)^2 \left(\left\langle S_p \right\rangle - \left\langle S_n \right\rangle \right)^2.\label{eqn:sm_defs}
\end{eqnarray}
Here, $\mathcal{Z}$ and $\mathcal{N}$ are the proton and neutron numbers, respectively, $\Delta^{n/p}_q$ \cite{ParticleDataGroup:2024cfk,Markisch:2018ndu,HERMES:2006jyl} are the hadronic matrix elements that parametrize the contribution of quark q to the spin content of a nucleon (n/p). The neutron and proton spin expectation values are represented by $\left\langle S_{p/n} \right\rangle$ \cite{Klos:2013rwa}. $Q^p_W$ and $Q^n_W$ are the Weak charges of the proton and neutron, respectively, defined in the SM to be,
\begin{eqnarray}
    Q^p_W &=& 2 C_u^V + C_d^V = 1 - 4 \sin^2\theta_W,\nn
    Q^n_W &=& C_u^V + 2 C_d^V = -1, \label{eqn:weak_charges}
\end{eqnarray}
where $C_q^V$ are the vector couplings of the quark q to the neutrino current and $\theta_W$ is the Weak mixing angle~\cite{ParticleDataGroup:2024cfk,Qweak:2018tjf}. The Axial form factor similarly depends on
\begin{eqnarray}
    g_A^0 = \frac12\left(g_A^p + g_A^n\right),\quad
    g_A^1 = \frac12\left(g_A^p - g_A^n\right),\quad\\
    g_A^N = \sum_{q=u,d,s} C_A^q\Delta^N_q, \quad N=p,n,
\end{eqnarray}
with $C_A^q$ being the axial couplings of the quark q to the neutrino current. All the couplings are assumed to be relative to the normalization factor $G_F/\sqrt{2}$, where $G_F$ is the Fermi constant.

The Weak and Axial form factors, $F_W(q^2)$ and $F_A(q^2)$, respectively, have been explicitly defined elsewhere~\cite{Hoferichter:2020osn,VanDessel:2020epd,Yang:2019pbx,Payne:2019wvy,Klos:2013rwa}, the former often being normalized to be unity at $q^2=0$. The zero-point value of the Weak form factor is conventionally expressed as the Weak charge of the nucleus, $Q_W$. It is to be noted, however, that $Q_W$ does not factor out of the expression for $F_W(q^2)$ and the respective nucleonic charges mix non-trivially for all non-vanishing transferred momenta.

Comparing the two zero-momentum values $\left\vert F_W(0)\right\vert^2$ and $F_A(0)$, one may infer the relative scales of the Vector and Axial contributions to $\cevns$ within the SM. Considering that the $\Delta^N_q$ are of the order of unity, and $\left[Q^p_W\sim 0.05\right]\ll \left[Q^n_W\sim -1\right]$, the Axial contribution is suppressed by at least a factor of $\sim \mathcal{N}^2$ relative to the Vector contribution. This is the origin of the "coherent enhancement" of the Vector contribution and the reason why $\cevns$ is generally considered a coherent process. The Axial contribution is, however, a non-negligible component that does not vanish as $E_R\rightarrow 0$ and may be quite relevant under certain circumstances, for example, for light, neutron-deficient nuclei with non-zero nuclear spin or for testing certain BSM scenarios. It may also be used to probe the nuclear spin structure functions, which are of interest in their own right, for example, in the context of Dark Matter searches.

The kinematic factors $\mathcal{K}_{V/A}$ are functions of the nuclear recoil energy $E_R$, the incident neutrino energy $E_\nu$ and the mass $M_N$ of the nucleus. They are both unity at $E_R=0$ and differ only at higher orders in $E_R/E_\nu$. Along with the Weak form factor, these factors largely dominate the shape of the $\cevns$ spectrum and, when combined with the respective nuclear responses, provide a unique window into the relative sensitivity of the spectra to the various microphysical and nuclear effects. The two components taken together capture overwhelmingly both the isotope-dependence and the $(E_R, E_\nu)$-dependence of the $\cevns$ spectrum. All other target and neutrino source dependencies, whether within the SM or beyond, are suppressed in comparison by several orders of magnitude, at least.
The interference between the Vector and Axial contributions, appearing at the next order and vanishing as $E_R\rightarrow 0$, has been neglected here.

Traditionally, parametric approximations to the explicit Weak form factor have been used for modeling the Weak nuclear response. The Weak charge is typically factored out and a mapping of $F_W(q^2) \rightarrow Q_W F_{\rm ref}(q^2)$ has always been implemented. Three different parameterizations are most popular for this purpose, viz. the Helm~\cite{Helm:1956zz}, the Symmetrized-Fermi~\cite{Sprung_1997} and the Klein-Nystrand~\cite{Klein:1999qj} form factors. For a detailed description, see, for example, Ref.~\cite{Lewin:1995rx,AristizabalSierra:2019zmy}. Each parametrization has its own set of free parameters, generally constrained by empirical inference of nuclear charge distributions from electron scattering and muonic-atom spectroscopy~\cite{DeVries:1987atn,Fricke:1995zz,Lewin:1995rx}, which are used to describe common nuclear parameters. The Weak radius of the nuclei ($R_W$) is in this case the most pertinent nuclear parameter that may be used as a global fit parameter to describe the Weak form factor. The respective predictions begin to disagree at high transferred momenta, which should be taken into account, especially for scattering off high-energy neutrinos (e.g., atmospheric neutrinos at $\gtrsim 100$ MeV). For neutrinos at $E_\nu \lesssim 50$ MeV, $\cevns$ predictions using the Helm or Klein-Nystrand parameterizations and first-principle evaluations using Chiral Effective Field Theory-Shell Model ($\chi$EFT-SM)~\cite{Hoferichter:2020osn} or Hartree-Fock approach~\cite{VanDessel:2020epd} have been shown to agree within $1\%$ for $^{40}$Ar nuclei~\cite{Carey:2025byg}. Including Relativistic Mean Field evaluations~\cite{Yang:2019pbx} and NNLO {\it ab initio} calculations of Ref.~\cite{Payne:2019wvy} extends the agreement to within $3\%$ in the same case.

\subsection*{Beyond the Standard Model}

Going beyond the SM, the Generalized Neutrino Interaction (GNI) framework described in Refs.~\cite{Lindner:2016wff,AristizabalSierra:2018eqm,Bischer:2018zcz} extends the popular neutrino NSI framework to include Scalar and Tensor currents. Ref.~\cite{Hoferichter:2020osn} further describes the relevant EFT at dimensions 5, 6 \& 7 through which $\cevns$ proceeds~\cite{Hoferichter:2018acd,Altmannshofer:2018xyo} and connects it explicitly to the nuclear response matrices that also affect Dark Matter (DM) searches via nuclear recoils~\cite{Fitzpatrick:2012ix,Anand:2013yka},
\begin{eqnarray}
    \mathcal{L}^{(5)} &=& \frac{G_F}{\sqrt{2}}\,C_F\Bar{\nu}\sigma^{\mu\nu}P_L\nu F_{\mu\nu},\nn
    \mathcal{L}^{(6)} &=& \frac{G_F}{\sqrt{2}}\left(C_q^V \Bar{\nu} \gamma^\mu P_L \nu \Bar{q}\gamma_\mu q 
    + C_q^A \Bar{\nu} \gamma^\mu P_L \nu \Bar{q}\gamma_\mu \gamma^5q\right)\nn
    &&+ \frac{G_F}{\sqrt{2}}\,C_q^T \Bar{\nu} \sigma^{\mu\nu} P_L \nu \Bar{q}\sigma^{\mu\nu}q \nn
    \mathcal{L}^{(7)} &=& \frac{G_F}{\sqrt{2}}\left[\left(C_q^S + \frac{8\pi}{9} C_g^{\prime S}\right) \Bar{\nu} P_L \nu m_q \Bar{q}q\right.\nn
    &&\left.+ C_q^P \Bar{\nu} P_L \nu m_q \Bar{q} i \gamma^5 q 
    - \frac{8\pi}{9} C_g^{\prime S} \Bar{\nu} P_L \nu \theta_\mu^\mu\right], \label{eqn:eft}
\end{eqnarray}
where $\theta_\mu^\mu$ is the QCD trace anomaly.
The neutrino indices have been suppressed since current $\cevns$ experiments are not sensitive to neutrino flavor-transitions. A sum over only the light quarks ($q=u,d,s$) in all combinations are generally implied, the heavier quarks having been integrated out and their effect absorbed into the scalar coupling $C_g^{\prime S}$. The projection operator $P_L$ excludes right-handed neutrinos from appearing explicitly, reducing the number of independent operators by half compared to the DM-nucleus EFT. Probing these excluded operators would need right-handed neutrino beams. However, the tensor and scalar neutrino bilinears do still require final-state right-handed neutrinos. Additionally, all flavor-diagonal vector and tensor couplings vanish identically for Majorana neutrinos. For further details, one may refer to Refs.~\cite{Hoferichter:2020osn, Hoferichter:2018acd, Altmannshofer:2018xyo,SHIFMAN1978443, Goodman:2010ku}.

Generalizing these prescriptions, the full $\cevns$ differential cross-section, including the BSM contribution, has been summarized as~\cite{Lindner:2016wff,AristizabalSierra:2018eqm,Bischer:2018zcz,Hoferichter:2020osn},
\begin{eqnarray}
    \left.\frac{d\sigma}{dE_R}\right\vert_{BSM} &=& \frac{G_F^2 M_N}{4\pi}\times[\mathcal{K}_V\, \left\vert F_W(q^2)\right\vert^2\nn
    &+& \mathcal{K}_A\, F_A(q^2) + \mathcal{K}_S\, \left\vert F_S(q^2)\right\vert^2\nn
    &+& 8\,\mathcal{K}_T^\mathcal{T}\, F_T^\mathcal{T}(q^2) + 16\,\mathcal{K}_T^\mathcal{L}\, F_T^\mathcal{L}(q^2) \nn
    &\pm& \mathcal{K}_{int}\, \left(F_{AV}(q^2) - R\right)]. \label{eqn:rate_bsm} 
\end{eqnarray}
Where the new kinematic factors are,
\begin{eqnarray}
    \mathcal{K}_S &=& \frac{M_N E_R}{E_\nu^2} + \frac{E_R^2}{2E_\nu^2}\nn
    \mathcal{K}_T^\mathcal{T} &=& 1 - \frac{M_N E_R}{2E_\nu^2} -\frac{E_R}{E_\nu} + \frac{E_R^2}{2E_\nu^2}\nn
    \mathcal{K}_T^\mathcal{L} &=& 1-\frac{E_R}{E_\nu}\nn
    \mathcal{K}_{int} &=& \frac{E_R}{E_\nu} - \frac{E_R^2}{E_\nu^2}. \label{eqn:bsm_kinem}
\end{eqnarray}
It is to be stressed that, like in the SM, the quark-level couplings of Eqn.~\ref{eqn:eft} do not in general factor out of the form factors $F_x(q^2)$, but appear folded-in with their corresponding hadronic matrix elements. The full dependence on the underlying quark couplings is absorbed into the form factors themselves here, rather than introducing separate overall coefficients. The exact expressions of the form factors, their momentum dependence, and the values of the hadronic matrix elements are not directly relevant for this work. Detailed descriptions may be found in, for example, Refs.~\cite{Hoferichter:2020osn,Markisch:2018ndu,HERMES:2006jyl,Hoferichter:2018acd,schneider:2017Sc,CODATA:2025,antognini:2013,Koester:1995,Kopecky:1997,Djukanovic:2019,Alexandrou:2020,Hill_2018,Gupta:2018,Hoferichter:2019,Hoferichter:2015rse,Durr:2016lat,Yang:2016chf,Abdel-Rehim:2016etm,Bali:2016lqm,Hill:2015whq,Hoferichter:2012rsp,Hoferichter:2012dit,Fodor:2016udq}. Of the BSM contributions, only the vector and scalar components are coherently enhanced. The scalar contribution, however, is only relevant in the presence of right-handed neutrinos and vanishes as $q^2\rightarrow0$. Unlike the vector and scalar BSM components, the rest of the components are again suppressed by a factor $\sim \mathcal{N}^2$ and vanish identically for $J_N=0$. 

The effect of the pseudoscalar current arising at dimension-7 in Eqn.~\ref{eqn:eft} is usually absorbed within the spin-dependent structure factors, with a similar $\sim \mathcal{N}^2$ suppression and is thus not included explicitly.
In addition, they appear with a leading kinematic factor of $\sim E_\nu^2/M_N^2$ which, in the present regime, corresponds to approximately the same numerical scale as $E_R^2/E_\nu^2$ but with no SM counterparts. The dipole term at dimension-5 coupling to the EM field strength $F_{\mu\nu}$, comes with a leading kinematic factor of $1/M_N E_R$. 
This produces an additional contribution to Eqn.~\ref{eqn:rate_bsm} at a relative scale of $\lesssim 10^{-15}\mathcal{Z}^2(C_F/eV)^2$, impactful at low energies in very light elements. 
This is also a new long-range interaction coupled to electric charge and is thus severely constrained by astrophysical data~\cite{Giunti:2015nel}. 
It may interfere with the BSM scalar contribution to produce a term proportional to $1/E_\nu$ at the leading order, which adds a term of relative strength $\lesssim 10^{-7}\mathcal{Z}(C_F/eV)$.
Coupled with the constraints on $C_F$, these terms may also be neglected given the current precision of $\cevns$ experiments. 
All the quark couplings in Eqn.~\ref{eqn:eft} are to be read relative to the explicit normalization factor $G_F/\sqrt{2}$, while their $q^2$-dependent effects are embodied directly in the form factors $F_x(q^2)$.
Eqns.~\ref{eqn:sm-rate}, \ref{eqn:sm_defs}, \ref{eqn:rate_bsm} \& \ref{eqn:bsm_kinem} motivate the possibility of a limited set of terms in $(E_R, E_\nu, M_N)$ that may be used to describe the most general $\cevns$ spectrum across all targets and neutrino sources. The multipliers are combinations of the various form factors and couplings, which may be constrained by $\cevns$ data.

%% file: sections/framework_intro.tex
\input{figures/ffsq_vs_er_fig.tex}
The predictive framework generates the predicted counts in the detector observables from the underlying physics and the detector response, subject to the control parameters. The physics input is the neutrino flux and the $\cevns$ cross-section, which together yield a predicted recoil spectrum. This spectrum is then mapped to the detector observables through the response matrix, yielding a predicted signal. This prediction may then be compared directly with the observed counts in a likelihood analysis.

At the physical recoil level, the differential rate per
target nucleus is
\begin{equation}\label{eq:continuous_rate}
\frac{dR}{dE_r}(E_r)
= N_T
\int_{E_\nu^{\min}(E_r)}^{E_\nu^{\max}}
\frac{d\sigma(E_r,E_\nu)}{dE_r}\,\frac{d\Phi}{dE_\nu}(E_\nu)\,dE_\nu,
\end{equation}
where $E_\nu^{\max}$ is the end-point energy of the neutrino flux and $E_\nu^{\min}(E_r)$ is the recoil-threshold neutrino energy from Eqn.\ref{eqn:kinematic_cond}. $N_T$ denotes the number of target nuclei involved for the given exposure.

The full prediction pipeline therefore incorporates three mutually independent components, {\it viz.} the differential cross-section, the neutrino flux and the detector response. The following subsections describe the implementation of each component and their final combination.

%% file: figures/ffsq_vs_er_fig.tex
\begin{figure*}[t]
\includegraphics[width=\textwidth]{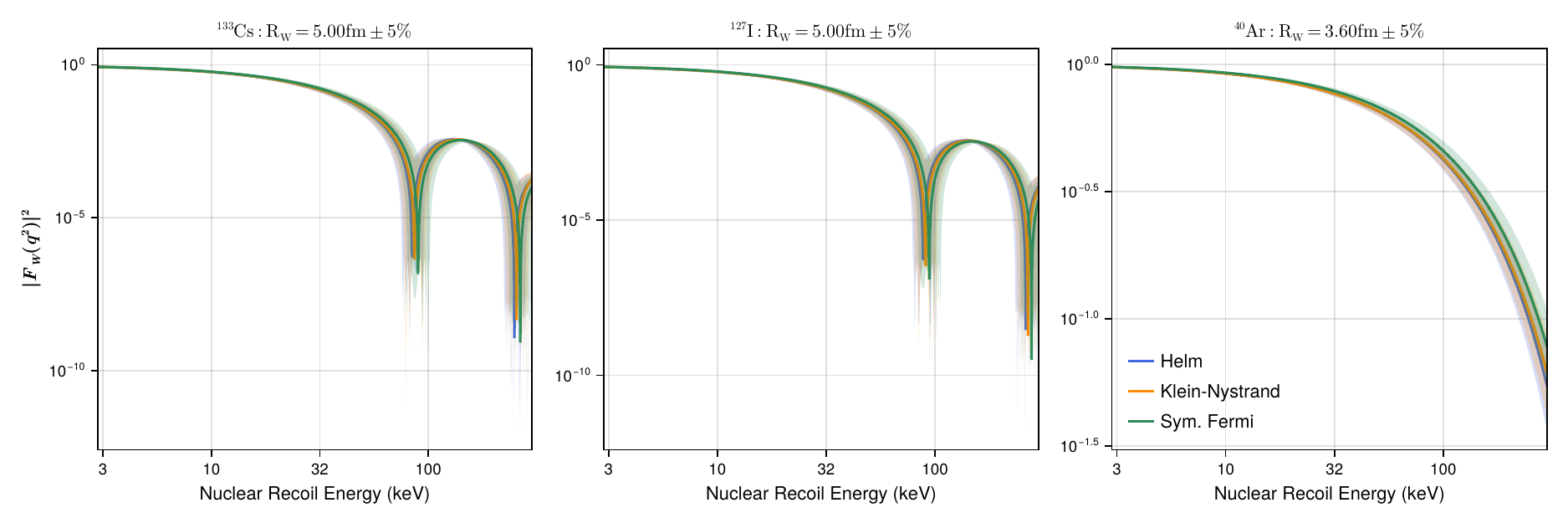}
\caption{\label{fig:ffsq_vs_er} Variation of the nuclear form factor squared $F^2(Q^2)$ with nuclear recoil energy $E_r$ for different nuclei. Choice of parameterizations are shown in different colors indicated in the legend. The shaded bands denote variation within the central 67\% interval of the Weak Radius $R_W$. The central values is shown as a solid line and the corresponding $R_W$ and isotope is indicated on top. }
\end{figure*}

%% file: sections/framework.tex
\input{figures/xsec_ar_fig}
\input{figures/xsec_csi_fig}
Irrespective of the origin and interpretation of each of the terms in Eqn.~\ref{eqn:rate_bsm}, whether within or beyond SM, the dominant $(E_R, E_\nu)-$dependent shape of the $\cevns$ differential cross-section depends uniquely on the dimensionless kinematic basis,
\begin{equation}
    \vec{\mathcal{B}} = \left\{ 1, \,\frac{M_N E_R}{E_\nu^2},\, \frac{E_R}{E_\nu},\, \frac{E_R^2}{E_\nu^2}\right\}\label{eqn:base}
\end{equation}
up to the second sub-leading order. Their relative scales are,
\begin{equation}
    \frac{M_N E_R}{E_\nu^2} \sim 1,\quad \frac{E_R}{E_\nu}\sim 10^{-3},\quad \frac{E_R^2}{E_\nu^2}\sim 10^{-6}\label{eqn:base_scaling}
\end{equation}
for the range of values of $M_N$, $E_R$ \& $E_\nu$ relevant to $\cevns$. Taking this into account, the most general $\cevns$ differential cross-section may now be written down as
\begin{eqnarray}
    \frac{d\sigma}{dE_R} = \frac{G_F^2 M_N}{4\pi} Q_{W,\rm SM}^2\left\vert F_{\rm ref}(q^2)\right\vert^2\vec{\mathcal{B}}\cdot\vec{\mathcal{P}}.\label{eqn:rate-fit}
\end{eqnarray}
with,
\begin{eqnarray}
    \vec{\mathcal{P}} &=& \Bigg\{\xi_1 + \xi_2 + \xi_4 + \xi_5,\nn
    &&-\frac12\left(\xi_1 - \xi_2 - 2\xi_3 + \xi_4\right),\nn
    &&-\left(\xi_1 + \xi_2 + \xi_4 + \xi_5 \mp \xi_1\xi_2\right),\nn
    &&\xi_1 + \xi_2 + \frac12\xi_3 + \frac12\xi_4 \mp \xi_1\xi_2\Bigg\},\label{eqn:p_decompose}
\end{eqnarray}
where,
\begin{eqnarray}
    &\xi_1 = \frac{\left\vert F_W(q^2)\right\vert^2}{Q_{W,\rm SM}^2\left\vert F_{\rm ref}(q^2)\right\vert^2},\quad
    \xi_2 = \frac{F_A(q^2)}{Q_{W,\rm SM}^2\left\vert F_{\rm ref}(q^2)\right\vert^2},\nn
    &\xi_3 = \frac{\left\vert F_S(q^2)\right\vert^2}{Q_{W,\rm SM}^2\left\vert F_{\rm ref}(q^2)\right\vert^2},\quad
    \xi_4 = \frac{\frac{64\pi}{2J_N+1} F_T^\mathcal{T}(q^2)}{Q_{W,\rm SM}^2\left\vert F_{\rm ref}(q^2)\right\vert^2},\nn
    &\xi_5 = \frac{\frac{128\pi}{2J_N+1} F_T^\mathcal{L}(q^2)}{Q_{W,\rm SM}^2\left\vert F_{\rm ref}(q^2)\right\vert^2}.\label{eqn:xi_decompose}
\end{eqnarray}
In the SM limit, only $\xi_1$ and $\xi_2$ are non-zero. In this limit, therefore, we have
\begin{eqnarray}
    \mathcal{P}_1 &=& \xi_1 + \xi_2,\nn
    \mathcal{P}_2 &=& -\frac12\left(\xi_1 - \xi_2\right),\nn
    \mathcal{P}_3 &=& -\left(\xi_1 + \xi_2 \mp \xi_1\xi_2\right),\nn
    \mathcal{P}_4 &=& \xi_1 + \xi_2 \mp \xi_1\xi_2.
\end{eqnarray}

For each target isotope, the cross-section module may be initialized with any one of the three reference parameterizations, viz. the Helm, Klein-Nystrand, or Symmetrized-Fermi profiles for $F_{\rm ref}(q^2)$. The variation of the different form factors over the relevant range of nuclear recoil energies, as used in this work, is shown in Fig.~\ref{fig:ffsq_vs_er}. The Helm implementation is produced by~\cite{Helm:1956zz},
\begin{eqnarray}
    F_{\rm ref}^{\rm Helm}(q^2) &=& 3\frac{j_1(qR_0)}{qR_0}e^{-(qs)^2/2},\nn
    q^2 &=& 2M_N E_R,\nn
    R_0^2 &=& \frac53\left(R_W^2 - 3s^2\right),\nn
    s &=& 0.9\,{\rm fm}.
\end{eqnarray} 
The Klein-Nystrand implementation is produced by\cite{Klein:1999qj},
\begin{eqnarray}
    F_{\rm ref}^{\rm KN}(q^2) &=& 3\frac{j_1(qR_A)}{qR_A}\frac{1}{1+(qa)^2},\nn
    q^2 &=& 2 M_N E_R,\nn
    R_A^2 &=& \frac{5}{3}\left(R_W^2 - 6a^2\right),\nn
    a &=& 0.7\,{\rm fm}.
\end{eqnarray}
The Symmetrized-Fermi implementation is produced by~\cite{Sprung_1997},
\begin{eqnarray}
    F_{\rm ref}^{\rm SF}(q^2) &=& \frac{3}{q c\left((qc)^2 + (\pi qa)^2\right)}\left[\frac{\pi qa}{\sinh(\pi qa)}\right]\nn
    &\times&\left[\frac{\pi qa\sin(qc)}{\tanh(\pi qa)} - qc\cos(qc)\right],\nn
    q^2 &=& 2M_N E_R,\nn
    c &=& \sqrt{R_W^2 - \frac{7}{3}(\pi a)^2},\nn
    a &=& 0.7\,{\rm fm}.
\end{eqnarray}
The Weak radius $R_W$ is a property of the nucleus itself and expected, hence, to be independent of the choice of parameterization. Considering, $R_W$ is chosen to be the common nuisance parameter while $s$ and $a$ are fixed at their conventionally used values. The Weak charge ($Q_{W,\rm SM}$) is evaluated with a floating weak-mixing parameter centered on $\sin^2\theta_W=0.2382\pm 0.0011$~\cite{Qweak:2018tjf}, taken from its lowest energy determination besides $\cevns$.

In the SM limit one recovers $\vec{\mathcal{P}}=\vec{\mathcal{P}}_{\rm SM}=\left\{1,-\frac12,-1,1\right\}$ for scalar nuclei. For nuclei with non-zero spin, it also converges on this, albeit with a highly correlated $J_N-$dependent correction, suppressed by a factor of approximately $\sim \mathcal{N}^2$. For the target elements currently used at $\cevns$ experiments and their reported resolutions, inferences on $\vec{\mathcal{P}}$ are not expected to be sensitive to these corrections. The dominant target dependence is factored out in the form of the weak charge $Q_{W,\rm SM}$ and form factor $F_{\rm ref}$. The remaining target dependence is largely contained in the basis set $\vec{\mathcal{B}}$ itself.
Further variation across targets from sub-leading form factors $F_A$, $F_S$ and $F_T^{\mathcal{T},\mathcal{L}}$ are expected to be correspondingly small relative to the dominant $F_W$ term.
At the level of Eqn.~\ref{eqn:rate-fit}, the dimensionless parameter set $\vec{\mathcal{P}}$ therefore captures the complete set of shape modifiers relative to the dominant Weak spectral profile, for whichever parameterization of $F_{\rm ref}(q^2)$ is chosen. From a purely analytic point of view, this is the most general, informed, polynomial shape that may be fit to available $\cevns$ spectral data, accurate up to a precision of $\mathcal{O}(10^{-6})$ relative to the dominant $\mathcal{O}(1)$ terms. Eqns.~\ref{eqn:p_decompose}~\&~\ref{eqn:xi_decompose} add a layer of physical interpretation to these parameters which are not necessary for the primary inference exercise. Given a specific choice of physical interpretation, the parameters may not be expected to remain independent, a fact that can be exploited to test the underlying hypothesis. 

The impact of variation in the components of $\vec{\mathcal{P}}$ on the differential cross-section for $^{40}$Ar and $^{133}$Cs are shown in Fig.~\ref{fig:xsec_ar} and Fig.~\ref{fig:xsec_csi}, respectively. (The impact of the components' variation for Iodine looks comparable to that of Cesium, and hence we omit it here.) The component $\mathcal{P}_1$, the most dominant parameter, has a strong impact on the overall normalization of the differential cross-section. The effect is typically independent of nuclear recoil and neutrino energies. The Weak form factor, on the other hand, rapidly falls at higher transferred momenta, which corresponds to higher $E_R$. Higher $E_{\nu}$ spreads the spectrum to higher $E_R$. At the same time, larger $E_R$ bins lead to a progressively shrinking window of integration over neutrino energies because of the kinematic constraint in Eqn.~\ref{eqn:kinematic_cond}, shifted in the direction of the highest neutrino energies. This high-tail kinematic suppression of the spectrum is in addition to the suppression due to loss of coherence, and is connected to the hard high-$E_{\nu}$ cutoff in the neutrino flux. This leads to a damping of $\mathcal{P}_1$ at high $E_R$ for high $E_{\nu}$.
The parameter $\mathcal{P}_2$ is modulated through the ratio $M_N E_R/E_{\nu}^2$. Considering that, for $\cevns$, $E_R$ is typically three orders of magnitude smaller than $E_{\nu}$, $\mathcal{P}_2$ has the strongest impact at mid- and high-$E_R$ for all neutrino energies. It is also strongly damped at high neutrino energies because of its suppression by $E_{\nu}^{-2}$, in addition to the loss of coherence.
The parameters $\mathcal{P}_3$ and $\mathcal{P}_4$, which are modulated through first and second powers of $E_R/E_{\nu}$, are considerably sub-dominant in comparison to the first two components. However, they do have a significant impact on the signal shape in their high tails. $\mathcal{P}_4$ affects even higher on in recoil energies than $\mathcal{P}_3$. The loss of coherence at high transferred momenta takes on a major role in this case. The onset of the decoherence regime depends on the target nucleus, with smaller, lighter nuclei remaining coherent until relatively higher recoil energies. Hence, smaller nuclei are expected to be more sensitive to these parameters. The kinematic suppression at high recoil energies also indicates that a neutrino flux with a larger high-$E_{\nu}$ population would be more appropriate for probing these parameters.

The cross-section sector contains two classes of uncertain quantities. The first are the basis coefficients themselves, which are assigned broad, uniform priors so that departures from the SM shape are driven primarily by the data rather than by any theory assumptions. The second are the parameters governing the SM weak charge and the reference form factor, viz. $\sin^2\theta_W$ and the isotope-dependent $R_W$. These are assigned informed priors based on the current state of knowledge, avoiding any determination that exploits $\cevns$ data itself.

The Weak mixing angle $\sin^2\theta_W$ runs with the renormalization scale, which is the transferred momentum ($q^2$) in practice. Theoretical evaluations based on current experimental measurements indicate that it varies by up to 3\% between the $Z$-pole and $246^{+54}_{-57}$MeV but remains relatively unchanged from there down to the Thomson limit~\cite{Erler:2017knj}. Experimental determinations below this threshold from M\o{}ller scattering ($\sin^2\theta_W=0.2397(10)(8)$ at $q\sim161$MeV)~\cite{Anthony:2005prl}, atomic parity violation on Caesium ($\sin^2\theta_W=0.2356(20)$ at $q=0.2$MeV)~\cite{Dzuba-apv:2012prl,Porsev-apv:2009prl}, and $e-p$ scattering ($\sin^2\theta_W=0.2382(11)$ at $q=158$MeV)~\cite{Qweak:2018tjf} agree with this within their respective experimental and theoretical uncertainties. Noting that, for nuclear recoil energies of up to $\sim 200$keV, we are dealing with $q\lesssim 230$MeV in $\cevns$, we adopt the latest and most precise measurement from the QWEAK Collaboration~\cite{Qweak:2018tjf}. The uncertainty on the theoretical evaluations is negligible compared to those from experimental determinations.

Due to the asymmetry of the neutron and proton Weak charges, the Weak Radius $R_W$ of the nucleus is largely a property of the neutron distribution of the nucleus. Experimental probes come primarily from Parity-violating Electron Scattering (PVES)~\cite{PhysRevLett.108.112502,PhysRevC.85.032501}. Theoretical evaluations utilize various methods and models which mostly agree with each other, within their respective uncertainties, where available~\cite{PhysRevC.100.054301,VanDessel:2020epd,PhysRevC.100.061304,Piekarewicz:2025lel,Hoferichter:2020osn}. Considering, the prior for $R_W$ sets a 10\% uncertainty on it for each isotope. This is larger than individual estimations but covers the spread among them.

Operationally, the differential cross-section matrices are not used in isolation. For each experiment, they are generated on the same $E_R$ and $E_\nu$ grids used elsewhere in the pipeline, then folded with the neutrino-flux arrays described in Subsec.~\ref{sec:neutrino_flux}. The signal prediction thus emerges in three successive steps: generation of isotope-resolved $d\sigma/dE_R$, folding with the experiment-specific flux in $E_\nu$ and time, and finally propagation through the detector response in recoil, timing, and observable space. This preserves the interpretation of $\vec{\mathcal{P}}$ as a shape descriptor at the cross-section level, while ensuring that all detector- and source-dependent effects are consistently accounted for only at the later stages of the likelihood construction.

The components of $\vec{\mathcal{P}}$ may, during inference, be correlated to each other at an algebraic level, depending upon the range of nuclear recoil and incident neutrino energies over which they are inferred. A broader effective window of inference over these energies would lead to a generally lower degree of correlation between the parameters. This effective window is determined by the overlap among the target recoil spectrum, the energy profile of the neutrino flux and the detector response.  The degree of correlation between specific parameters is also affected by the level of sensitivity of a particular dataset to the respective components of $\vec{\mathcal{P}}$. Although the degree of such correlations is specific to each dataset, their nature remains approximately the same across all targets, flux configurations, and detector sensitivities. 

This treatment also opens the possibility of a unified treatment of different target materials in a global analysis of $\cevns$ data, because the same $\vec{\mathcal{P}}$ description may be carried over to different targets at leading order. Significantly, a global coincidence of $\vec{\mathcal{P}}=\vec{\mathcal{P}}_{\rm SM}$ across isotopes is expected in the SM limit, and any observed deviation from this coincidence may be interpreted as a potential signal of unaccounted physics. The $\vec{\mathcal{P}}$ parameterization therefore provides a powerful tool for probing the underlying physics of $\cevns$ and for testing the SM in a model-independent manner. It also facilitates straightforward reinterpretation, since one may use the resulting constraints on $\vec{\mathcal{P}}$ to infer corresponding constraints on the underlying microphysical couplings and nuclear parameters once a specific realization of the form factors is adopted. Since the constraints themselves are relative to the choice of $F_{\rm ref}(q^2)$, they may also be used to test the validity of any specific model or to compare different models. The framework clearly segregates the multiple determinants of the $\cevns$ differential cross-section into domains of sensitivity described by the objects $(E_R, E_\nu, \mathcal{A}, J_N)$. Notably, not only do the inputs at the quark-, nucleonic-, and nuclear levels appear mixed in the specific response functions; these responses themselves are mixed within each of these domains of sensitivity. This formalizes the notion that multiple experiments, each sensitive to different combinations of these inputs, are required to disentangle the underlying physics. It provides a prescription for establishing complementary choices of target, neutrino source, and detector specifics to maximize sensitivity to specific classes of the multiple layers of physics that produce $\cevns$.

%% file: figures/xsec_ar_fig.tex
\begin{figure*}[t]
\includegraphics[width=\textwidth]{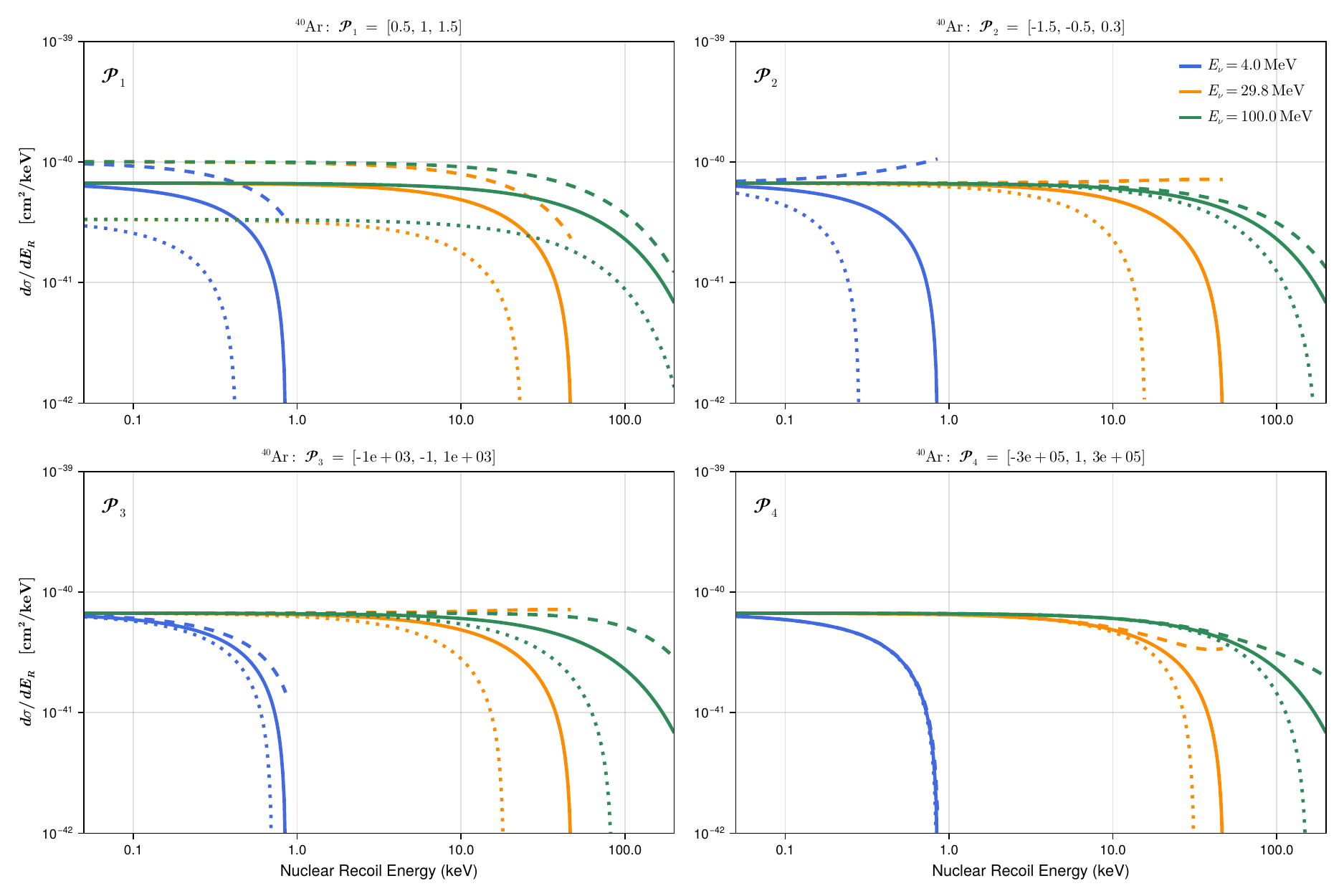}
\caption{\label{fig:xsec_ar} Impact of variation in components of the fit parameter $\vec{\mathcal{P}}$ on the $\cevns$ differential cross-section on $^{40}$Ar, as indicated in the upper left corner of each figure. Lines for varying neutrino energies are shown in different colors as indicated in the legend. Upward and downward variations within the central 67\% interval of the parameter range are shown as dashed and dotted lines, respectively. The range of the concerned parameter is indicated on top of each figure.}
\end{figure*}

%% file: figures/xsec_csi_fig.tex
\begin{figure*}[t]
\includegraphics[width=\textwidth]{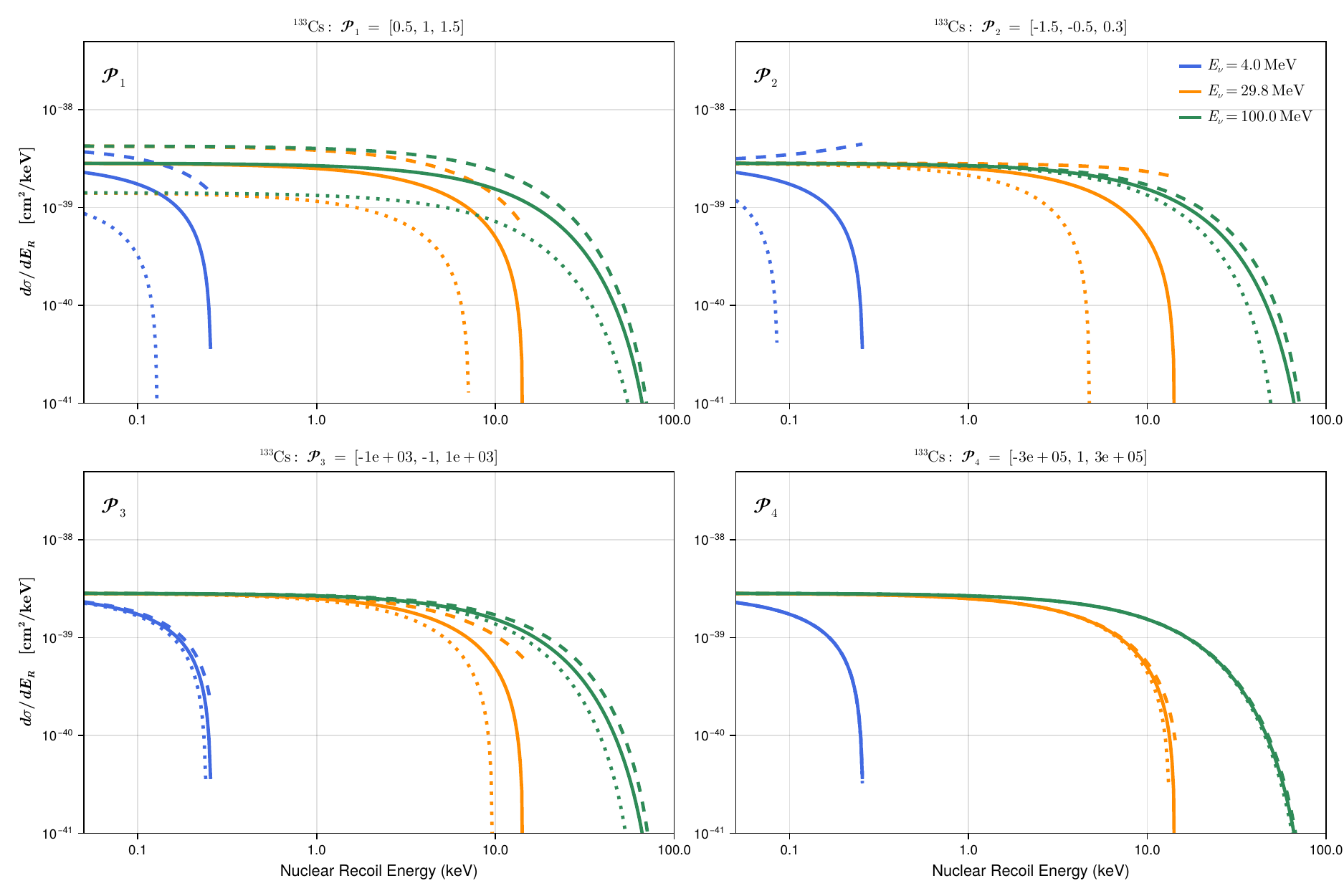}
\caption{\label{fig:xsec_csi} Impact of variation in components of the fit parameter $\vec{\mathcal{P}}$ on the $\cevns$ differential cross-section on $^{133}$Cs, as indicated in the upper left corner of each figure. Lines for varying neutrino energies are shown in different colors as indicated in the legend. Upward and downward variations within the central 67\% interval of the parameter range are shown as dashed and dotted lines, respectively. The range of the concerned parameter is indocated on top of each figure.}
\end{figure*}

%% file: figures/flux_envelope_fig.tex
\begin{figure*}[t]
\includegraphics[width=\textwidth]{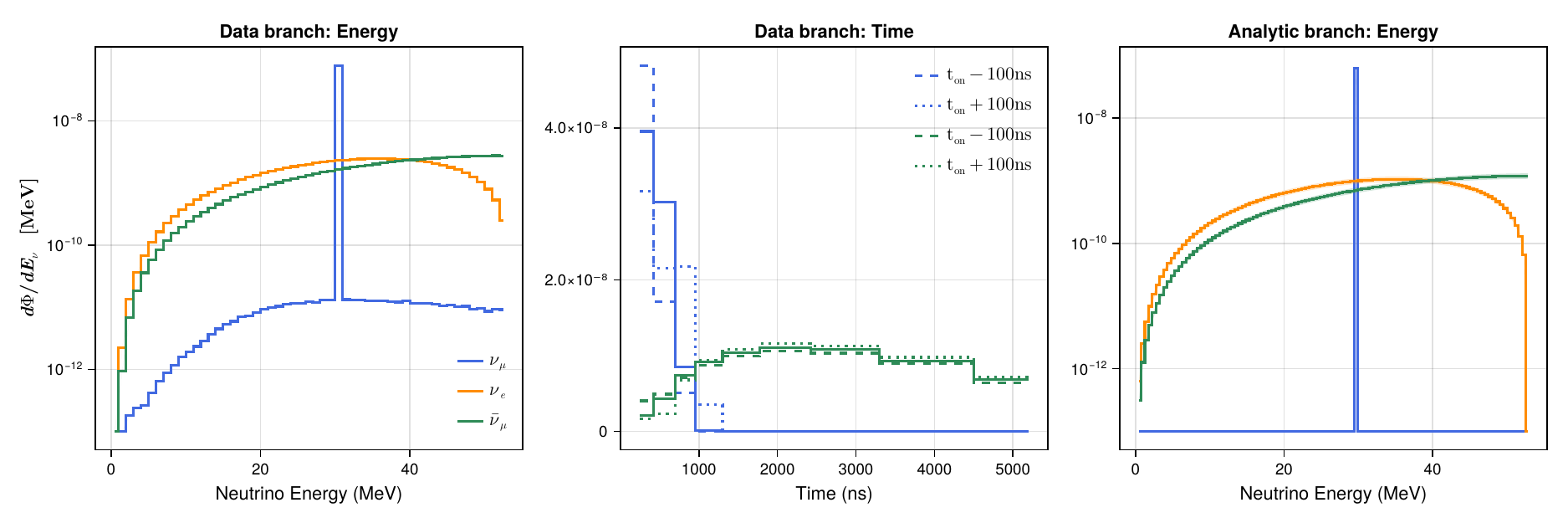}
\caption{\label{fig:flux_envelope} Neutrino flux prediction for COHERENT l-Ar \cite{COHERENT:2020iec} and CsI \cite{COHERENT:2021xmm} datasets. The data-driven branches use data provided by the COHERENT Collaboration in Ref.\cite{COHERENT:2021xmm}. The original analytic functions use the expressions of the Michel spectrum\cite{Michel:1949qe}. The fluxes are shown as a function of neutrino energy $E_\nu$ and time $t$ after proton on target (POT).}
\end{figure*}

%% file: sections/neutrino_flux.tex
Current dedicated $\cevns$ experiments~\cite{Akimov:2017ade,CCM:2021leg,NUCLEUS:2022zti,CONNIE:2024pwt,Ackermann:2025obx} are driven either by stopped-pion spallation neutrinos or by reactor antineutrinos. The stopped-pion neutrino flux is characterized by a prompt monoenergetic $\nu_\mu$ line at $E_\pi = (m_\pi^2-m_\mu^2)/(2m_\pi)$, followed by delayed $\nu_e$ and $\bar\nu_\mu$ components from muon decay at rest. The reactor antineutrino flux is characterized by a continuous spectrum of $\bar\nu_e$ with energies up to a few MeV.
The $\cevns$ datasets considered here \cite{COHERENT:2021xmm,COHERENT:2020iec,Akimov:2017ade} are driven by stopped-pion neutrinos at the SNS \cite{COHERENT:2021yvp}. The analytic forms of the SNS flux are well known \cite{Michel:1949qe}, and the COHERENT collaboration has also provided tabulated fluxes at the CsI detector location \cite{COHERENT:2021xmm}. The present framework is designed to accommodate either approach. The two branches are referred to as ``Data'' and ``Analytic'' in the following discussion.

The flux is treated as a matrix in $(E_\nu,t,\ldots)$ rather than as a one-dimensional spectrum in $E_\nu$ alone. This choice is physically motivated by the prompt/delayed structure of the source and is methodologically useful because it allows the same likelihood machinery to consistently propagate both spectral and timing information. Secondly, extensibility in dimensions is built in from the outset: the present implementation can accommodate higher-dimensional flux descriptors (for example, additional source, angular, or systematic coordinates) without altering the structure of the detector-response or cross-section folding steps. Only the $E_\nu$ axis is folded over while the rest are inherited by the flux-folded rate matrix. The CE$\nu$NS forward model consumes a standardized flux object, independent of whether that object is generated from analytic functional forms or from externally tabulated data. As a result, all experiments in the analysis can be treated within a common interface, with experiment-specific geometry and exposure entering only at the detector response level, or as transparent arguments passed at configuration. 

In the data-driven branch, we use the publicly available SNS flux tables \cite{COHERENT:2021xmm} (separate $\nu_\mu$, $\nu_e$, $\bar\nu_\mu$, $\bar\nu_e$ components), provided at the CsI detector location in units of MeV$^{-1}$ s$^{-1}$ cm$^{-2}$. Since those tables are detector-location fluxes, they are first unfolded to source-normalized fluxes by multiplying by
\begin{equation}
    4\pi L_{\rm CsI,pub}^2,\qquad L_{\rm CsI,pub}=1960\,{\rm cm}.
\end{equation}
The uncertainty on this distance is negligible for the present purpose. The source-normalized flux is then primed for each experiment through geometric dilution and beam-time scaling,
\begin{equation}
    \eta_{\rm data}(L,X,P_{\rm beam}) = \frac{1}{4\pi L^2}\,X\,\frac{3.6\times 10^6}{P_{\rm beam}},
\end{equation}
with $L$ in cm, exposure $X$ in GWh, and beam power $P_{\rm beam}$ in MW. Two nuisance parameters modulate the flux in this branch: a global flux normalization $f_{\rm norm}$ and an onset-time offset $\delta t_{\rm flux}$.

In the functional branch, the SNS flux is generated from the standard stopped-pion forms: a prompt monoenergetic $\nu_\mu$ line at
\begin{equation}
    E_\pi = \frac{m_\pi^2-m_\mu^2}{2m_\pi},
\end{equation}
and delayed components from muon decay at rest,
\begin{eqnarray}
    \phi_{\nu_e}(E_\nu) &\propto& 192\,\frac{E_\nu^2}{m_\mu^3}\left(\frac12-\frac{E_\nu}{m_\mu}\right),\nn
    \phi_{\bar\nu_\mu}(E_\nu) &\propto& 64\,\frac{E_\nu^2}{m_\mu^3}\left(\frac34-\frac{E_\nu}{m_\mu}\right).
\end{eqnarray}
with normalization written explicitly as
\begin{equation}
    \eta_{\rm func} = \frac{r\,N_{\rm POT}}{4\pi L^2},
\end{equation}
where $r$ is the neutrinos-per-flavor per POT and $N_{\rm POT}$ is obtained from beam energetics and exposure. Equivalently, with $X$ in GWh and proton-beam energy $E_p$,
\begin{equation}
    N_{\rm POT} = X\,\frac{3.6\times 10^{12}}{E_p\times 1.602\times 10^{-10}},
\end{equation}
so that $\eta_{\rm func}\propto X/(4\pi L^2)$. The parameter $r$ is the only nuisance parameter in this branch as it does not carry an inherent timing structure.
The resulting arrays are binned identically to the data-driven branch, for the respective axes, so that either branch can be used interchangeably in the same inference pipeline. The variation of the SNS flux with energy and time is shown in Fig.~\ref{fig:flux_envelope} for both the data-driven and functional branches. Additional dataset-specific details for the flux implementation and the nuisance parameters may be found in Apps.~\ref{app:lAr}~\&~\ref{app:CsI}.
Although the adopted flux is naturally connected to the datasets analyzed here, the framework does not require a specific choice. Any alternative flux prediction that can be mapped onto the same $(E_\nu,\ldots)$ representation can be plugged into the pipeline equivalently. This separation enables direct comparison of flux hypotheses without re-engineering the $\cevns$ analysis chain.
\input{figures/fit_spectrum}

%% file: figures/fit_spectrum.tex
\begin{figure*}[t]
    \centering

    \includegraphics[width=0.48\textwidth]{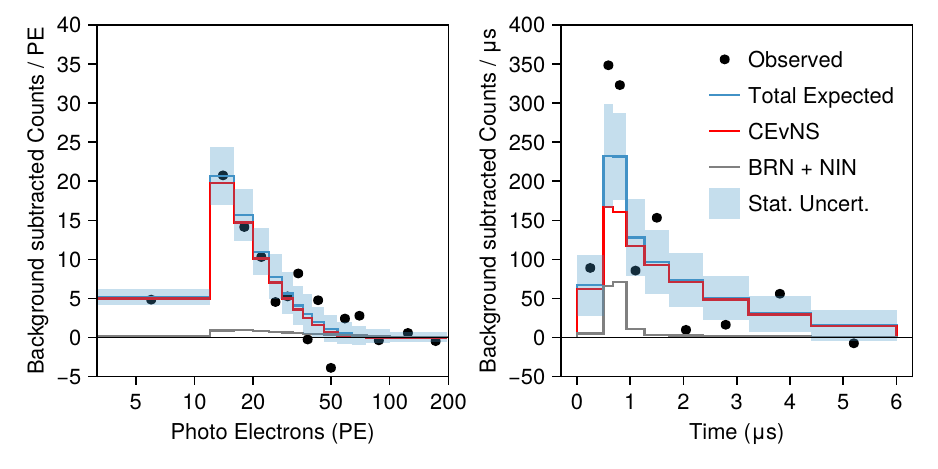}
    \hfill
    \includegraphics[width=0.48\textwidth]{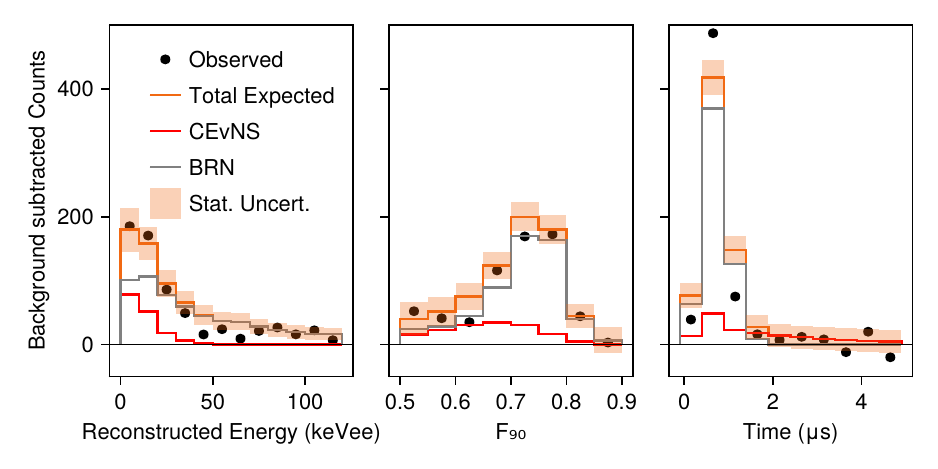}

    \caption{\label{fig:fit_spectrum}
    Best-fit spectrum projections on each of the observable axes, overlaid with the respective observed data, for the COHERENT CsI datasets (left) and the COHERENT l-Ar dataset (right), respectively.}
\end{figure*}

%% file: sections/response.tex
The flux-folded $\cevns$ rate from Eqn.\ref{eq:continuous_rate} is a prediction in the space of nuclear-recoil energy and, when timing information is retained, beam time. The experiments, however, observe a detector-dependent proxy, shaped by quenching, energy resolution, trigger and selection efficiencies, timing acceptance, and, in some cases, additional observables. For this reason, the conversion from theory space to data space is treated explicitly as a detector-response problem, following as closely as possible the detector prescriptions adopted in the corresponding experimental publications and data releases, which, in this case, include Refs.~\cite{COHERENT:2020iec,Akimov:2020czh,COHERENT:2021xmm}.

Detector effects are encoded in a response matrix
$\mathbf{A}\in\mathbb{R}^{n_{\mathrm{out}}\times n_E}$. The predicted counts are
\begin{equation}\label{eq:response_apply}
\mathbf{C}_{\mathrm{pred}} = \mathbf{A}\,\mathbf{R}_{\mathrm{tot}}.
\end{equation}
Here $A_{\alpha i}$ is the migration-and-acceptance weight from true recoil bin
$i$ to observed channel $\alpha$; row sums need not be unity in the presence of
efficiency loss or dead time.

The response is implemented on finite binning grids. If $\alpha,\beta,\ldots$ denote observable indices and $j,k,\ldots$ denote indices in the physical basis, one may write, in the most general factorized case,
\begin{equation}
	\mathbf{C}_{\mathrm{pred}}^{\alpha\beta\cdots} = \sum_{jk\cdots} A^{(1)}_{\alpha j} A^{(2)}_{\beta k}\cdots \mathbf{R}_{\mathrm{tot}}^{jk\cdots}.
\end{equation}
Equivalently, after vectorization of the corresponding tensors, the total detector operator may be written as a Kronecker product of the response matrices acting on uncorrelated observable axes,
\begin{equation}
	\mathbf{C}_{\mathrm{pred}} = \left(\mathbf{A}^{(1)}\otimes \mathbf{A}^{(2)}\otimes \cdots \right)\mathbf{R}_{\mathrm{tot}}.
\end{equation}
This makes the framework's expansivity explicit: additional observables can be incorporated by appending further response factors, provided that the corresponding detector-effect factors factorize to a good approximation in the physical basis. The uncertain detector parameters are therefore treated as part of the global inference problem rather than as fixed external corrections. Quenching prescriptions, efficiency deformations, fiducial masses, and background normalizations are all propagated through the forward model at the same level as the physics parameters. Future additional datasets can be incorporated by adding new detector-response modules with their own set of nuisance parameters alongside the observed data. This is essential for a global $\cevns$ analysis, because detector effects can mimic or absorb genuine spectral distortions if not consistently marginalized. 

The resulting framework retains a clean division of labor: flux and cross-section modules generate the physical rate, while the detector-response module translates that rate into the exact binned observables used by each dataset. Explicit details regarding the implementation of the two datasets considered in this work may be found in Apps.\ref{app:lAr}\&\ref{app:CsI}.

%% file: sections/fwd_model.tex
For a binned analysis, the exact bin integral of the rate is
\begin{eqnarray}\label{eq:bin_integral_exact}
R_i
&=& \int_{I_i}\frac{dR}{dE_r}(E_r)\,dE_r\nn
&=& N_T\!\int_{I_i}\!dE_r
\int_{E_\nu^{\min}(E_r)}^{E_\nu^{\max}}
\frac{d\sigma(E_r,E_\nu)}{dE_r}\,\frac{d\Phi}{dE_\nu}(E_\nu)\,dE_\nu,\nn
\end{eqnarray}
where $I_i$ denotes the recoil bin $i$ with center $E_{r,i}$ and
width $\Delta E_{r,i}$.
For sufficiently fine recoil binning, one obtains the standard approximation
\begin{equation}\label{eq:bin_approx}
R_i \approx
N_T\,\Delta E_{r,i}
\sum_{j=1}^{n_\nu}
\frac{d\sigma(E_{r,i},E_{\nu,j})}{dE_r}
\,\frac{d\Phi}{dE_\nu}(E_{\nu,j})\,\Delta E_{\nu,j}.
\end{equation}
Operationally, this evaluation is compacted as a matrix multiplication,
\begin{equation}\label{eq:matrix_fold}
\begin{aligned}
\mathbf{R}
&= N_T\,\mathrm{diag}(\Delta\mathbf{E}_r)\,\mathbf{D}
\,(\Delta\mathbf{E}_\nu\,\boldsymbol{\Phi}),
\end{aligned}
\end{equation}
using the following definitions,
\begin{eqnarray*}
&&D_{ij} \equiv \frac{d\sigma(E_{r,i},E_{\nu,j})}{dE_r},
\qquad
\Phi_j \equiv \frac{d\Phi}{dE_\nu}(E_{\nu,j}),\\
&&\Delta\mathbf{E}_\nu = \mathrm{diag}(\Delta E_{\nu,1},\ldots,\Delta E_{\nu,n_\nu})\\
&&\Delta\mathbf{E}_r=(\Delta E_{r,1},\ldots,\Delta E_{r,n_E})^\top.
\end{eqnarray*}

The differential cross-section and flux are evaluated on a grid of $n_E$ recoil bins and $n_\nu$ neutrino-energy bins. The bin widths $\Delta E_{r,i}$ and $\Delta E_{\nu,j}$ are included explicitly to ensure that the result is independent of the binning choice. The dimensions are mutable and are derived from the explicitly defined recoil and neutrino energy bins in the experimental dataset and neutrino flux modules, respectively. While the computation itself is agnostic to the choice of binning, it must be chosen to ensure that the approximation in Eq.~\eqref{eq:bin_approx} remains valid.
For a multi-isotope target with fractions $f_k$ ($\sum_k f_k=1$), each isotope
contributes its own $\mathbf{D}^{(k)}$, giving
\begin{equation}\label{eq:isotopes_sum}
\mathbf{R}_{\mathrm{tot}}
= \sum_k f_k\,N_T\,\mathrm{diag}(\Delta\mathbf{E}_r)
\,\mathbf{D}^{(k)}\,(\Delta\mathbf{E}_\nu\,\boldsymbol{\Phi}).
\end{equation}

The full pipeline can be summarized compactly as,
\begin{equation*}
\mathbf{D}^{(k)}(E_r,E_\nu)
\rightarrow
\mathbf{R}_{\mathrm{tot}}(E_r)
\rightarrow
\mathbf{C}_{\mathrm{pred}}(\mathrm{obs}),
\end{equation*}
with flux folding and bin-width factors applied in Eqs.~\eqref{eq:matrix_fold}
and \eqref{eq:isotopes_sum}.

The forward model does not have any fixed set of free parameters. It inherits the full set of parameters from the individual modules producing each of the components $\mathbf{D}^{(k)}$, $\boldsymbol{\Phi}$, and $\mathbf{A}$. These modules are quasi-independent and may be inspected, swapped out, or modified individually. To use the forward model with a dataset, each component must be initialized correctly for that dataset. The final analysis treats all parameters on equal footing. The posteriors on any set of parameters are obtained by marginalizing over the remaining parameters, which may be nuisance parameters or parameters of interest.
The accuracy of Eq.~\eqref{eq:bin_approx} is controlled by recoil-bin size
and integrand smoothness; higher-order quadrature can be used if required.
Non-uniform binning is naturally handled through $\Delta\mathbf{E}_\nu$ and
$\mathrm{diag}(\Delta\mathbf{E}_r)$. Apart from the neutrino energy dimension, all other dimensions in $\boldsymbol{\Phi}$ and $\mathbf{D}^{(k)}$ are inherited by $\mathbf{R}_{\mathrm{tot}}$. The detector response $\mathbf{A}$ may typically be designed to consume any combination of these dimensions into a set of observables.
The analysis uses a binwise Poisson likelihood to model the observed counts $\mathbf{C}_{\mathrm{obs}}$ given the predicted counts $\mathbf{C}_{\mathrm{pred}}$.

%% file: figures/corner_copy.tex
\begin{figure*}[t]
    \centering

    \begin{subfigure}
        \centering
        \includegraphics[width=0.8\textwidth]{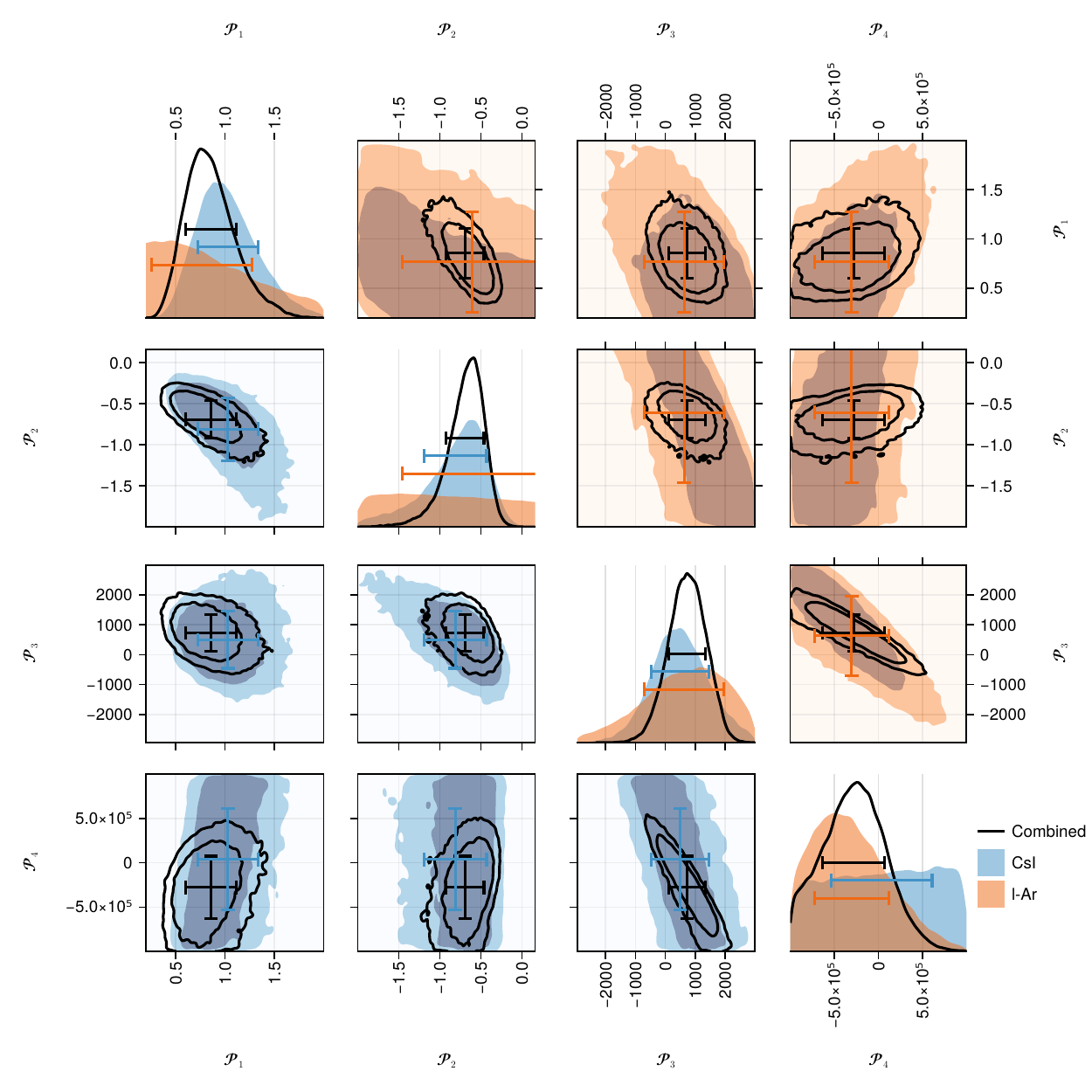}
    \end{subfigure}

    \begin{subfigure}
        \centering
        \includegraphics[width=0.8\textwidth]{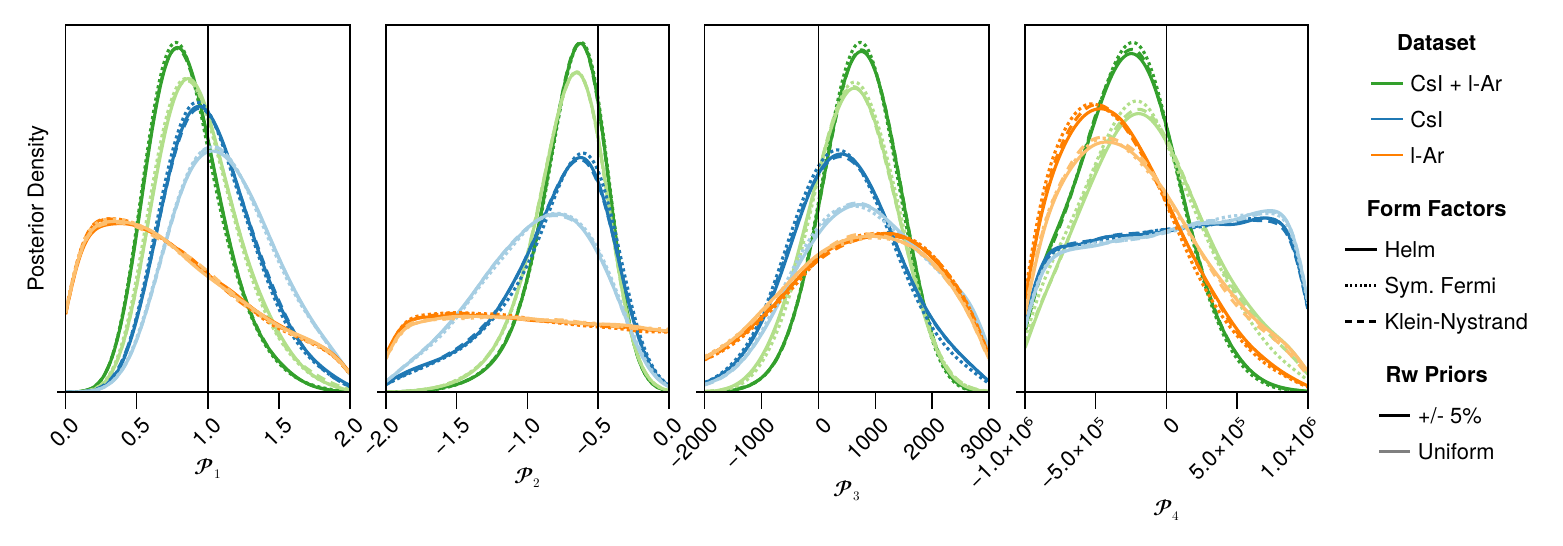}
    \end{subfigure}

    \caption{\label{fig:corner}
    Posteriors on the parameters of interest from the analysis of the COHERENT CsI and l-Ar datasets. The contours correspond to 68\% and 90\% credible intervals. The color coding corresponds to the individual analyses of the datasets, while the combined analysis is shown in black. The lower panel shows the corresponding marginalized posteriors for $\vec{\mathcal{P}}$ from each individual analysis and the global analysis.}
\end{figure*}

%% file: figures/cevns_xsec_fig.tex
\begin{figure*}[t]
\includegraphics[width=\textwidth]{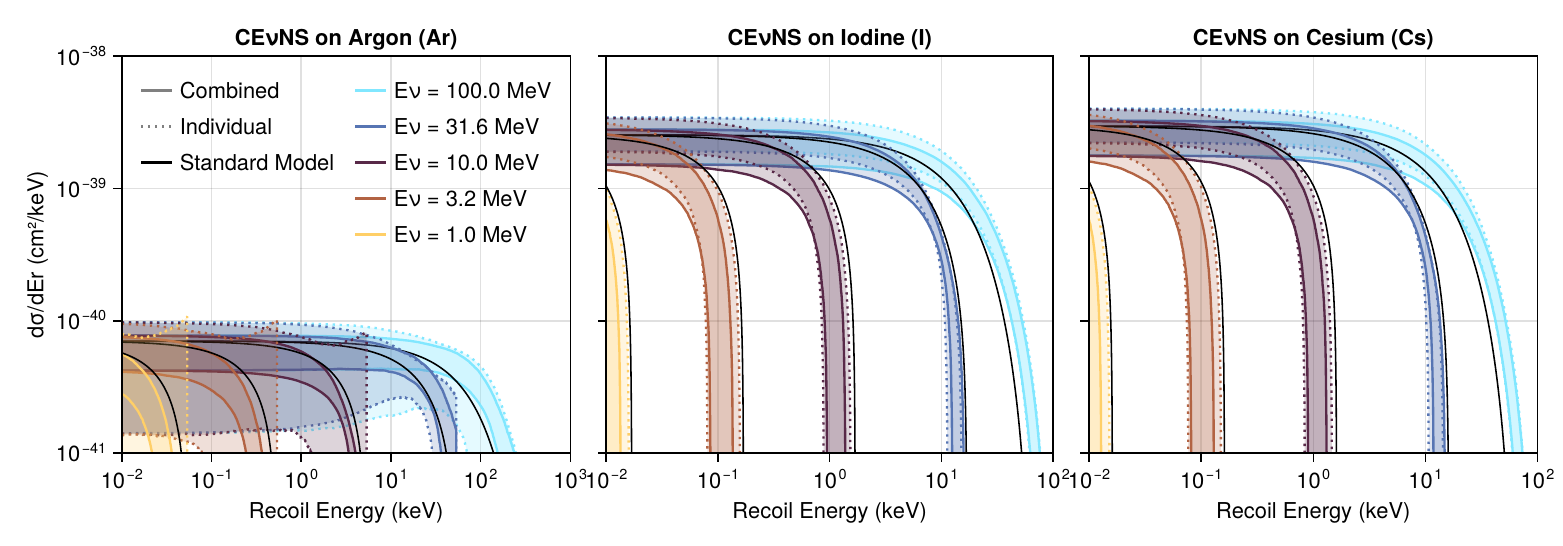}
\caption{\label{fig:cevns_xsec} The best-fit $\cevns$ differential cross-section on $^{40}\text{Ar}$, $^{133}\text{Cs}$ and $^{127}\text{I}$ as a function of nuclear recoil energy from the COHERENT CsI and Ar datasets. The shaded bands correspond to $68\%$ confidence intervals. The solid black and coloured lines represent the Standard Model prediction and the combined best-fit prediction, respectively. The dashed lines represent the best-fit predictions from the individual analyses of the COHERENT CsI and Ar datasets.}
\end{figure*}

%% file: tables/params.tex
\caption{List of parameters used in each of the analyses, along with their nominal values and prior distributions.}\label{tab:params}
\resizebox{\textheight}{!}{%
\begin{tabular}{|l|lll|lll|}
\hline
\textbf{Module} &
  \multicolumn{3}{c|}{\textbf{CsI{[}Na{]}}} &
  \multicolumn{3}{c|}{\textbf{l-Ar {[}CENNS10{]}}} \\ \hline
 &
  \multicolumn{1}{c|}{\textbf{Parameter}} &
  \multicolumn{1}{c|}{\textbf{Nominal}} &
  \multicolumn{1}{c|}{\textbf{Prior}} &
  \multicolumn{1}{c|}{\textbf{Parameter}} &
  \multicolumn{1}{c|}{\textbf{Nominal}} &
  \multicolumn{1}{c|}{\textbf{Prior}} \\ \hline
\multirow{3}{*}{Cross-section} &
  \multicolumn{1}{l|}{${\text R}_{\text W}$($^{133}{\text Cs}$, $^{127}{\text I}$)} &
  \multicolumn{1}{l|}{5.0 fm} &
  Uniform/Normal(10\%) &
  \multicolumn{1}{l|}{${\text R}_{\text W}$($^{40}{\text Ar}$)} &
  \multicolumn{1}{l|}{3.6 fm} &
  Uniform/Normal(10\%) \\
 &
  \multicolumn{1}{l|}{$\sin^2\theta_W$} &
  \multicolumn{1}{l|}{0.2382} &
  Normal(0.0011) &
  \multicolumn{1}{l|}{$\sin^2\theta_W$} &
  \multicolumn{1}{l|}{0.2382} &
  Normal(0.0011) \\
 &
  \multicolumn{1}{l|}{$\vec{\mathcal{P}}$} &
  \multicolumn{1}{l|}{\{1.0, -0.5, -1.0, 1.0\}} &
  Uniform &
  \multicolumn{1}{l|}{$\vec{\mathcal{P}}$} &
  \multicolumn{1}{l|}{\{1.0, -0.5, -1.0, 1.0\}} &
  Uniform \\ \hline
\multirow{2}{*}{Neutrino Flux} &
  \multicolumn{1}{l|}{Normalization ($f_{norm}$)} &
  \multicolumn{1}{l|}{1.0} &
  Normal(10\%) &
  \multicolumn{1}{l|}{Neutrinos per POT ($r$)} &
  \multicolumn{1}{l|}{0.09} &
  Normal(10\%) \\
 &
  \multicolumn{1}{l|}{Onset-time offset ($\delta t_{flux}$)} &
  \multicolumn{1}{l|}{0.0 ns} &
  Normal (100) &
  \multicolumn{1}{l|}{} &
  \multicolumn{1}{l|}{} &
   \\ \hline
\multirow{2}{*}{Detector Response} &
  \multicolumn{1}{l|}{Efficiency Mode ($z_\epsilon$)} &
  \multicolumn{1}{l|}{0.0} &
  Normal(1.0) &
  \multicolumn{1}{l|}{Fiducial Mass} &
  \multicolumn{1}{l|}{24.4 kg} &
  Normal (0.61) \\
 &
  \multicolumn{1}{l|}{QF modes ($z_0$, $z_1$)} &
  \multicolumn{1}{l|}{0.0} &
  Normal(1.0) &
  \multicolumn{1}{l|}{QF Modes ($a$, $b$)} &
  \multicolumn{1}{l|}{(0.246, 0.00078)} &
  Normal(0.006, 9.0e-05) \\ \hline
\multirow{3}{*}{Backgrounds} &
  \multicolumn{1}{l|}{BRN} &
  \multicolumn{1}{l|}{27.4} &
  Normal(6.9) &
  \multicolumn{1}{l|}{Prompt BRN} &
  \multicolumn{1}{l|}{497} &
  Normal(159) \\
 &
  \multicolumn{1}{l|}{NIN} &
  \multicolumn{1}{l|}{6.9} &
  Normal(2.5) &
  \multicolumn{1}{l|}{Delayed BRN} &
  \multicolumn{1}{l|}{33} &
  Normal(33) \\
 &
  \multicolumn{1}{l|}{SS} &
  \multicolumn{1}{l|}{3889} &
  Normal(81.7) &
  \multicolumn{1}{l|}{SS} &
  \multicolumn{1}{l|}{3152} &
  Normal(25) \\ \hline
\end{tabular}
}

%% file: tables/norm_fit_data.tex
\caption{Resulting best-fit values of the norms of the signal and background components. The values are given for the three nuclear form-factor models considered, together with the global best-fit estimate.}
\label{tab:norm_fit}\resizebox{\textheight}{!}{%
\begin{tabular}{|l|l|l|l|l|l|l|l|l|l|l|l|}
\hline
\multicolumn{6}{|c|}{\textbf{CsI{[}Na{]}}} &
\multicolumn{6}{c|}{\textbf{l-Ar (CENNS10)}} \\ \hline
\textbf{Component} & \textbf{Prior} & \textbf{Helm} & \textbf{Klein-Nystrand} & \textbf{Sym. Fermi} & \textbf{Global Estimate} &
\textbf{Component} & \textbf{Prior} & \textbf{Helm} & \textbf{Klein-Nystrand} & \textbf{Sym. Fermi} & \textbf{Global Estimate} \\ \hline

CE$\nu$NS & 364.7 $\pm$ 21.8 & 333.4 $\pm$ 36.4 & 356.0 $\pm$ 37.4 & 312.5 $\pm$ 35.3 & 328.9 $\pm$ 30.9 &
CE$\nu$NS & 125.9 $\pm$ 3.3 & 157.0 $\pm$ 54.9 & 188.1 $\pm$ 54.0 & 162.9 $\pm$ 52.0 & 156.0 $\pm$ 43.3 \\ \hline

BRN & 27.4 $\pm$ 6.7 & 31.6 $\pm$ 7.0 & 31.7 $\pm$ 6.5 & 33.2 $\pm$ 6.5 & 32.6 $\pm$ 6.6 &
BRN & 497.0 $\pm$ 160.9 & 538.6 $\pm$ 37.6 & 540.0 $\pm$ 36.9 & 560.7 $\pm$ 35.1 & 557.9 $\pm$ 33.7 \\ \hline

NIN & 6.9 $\pm$ 2.5 & 8.7 $\pm$ 2.4 & 6.3 $\pm$ 2.4 & 8.1 $\pm$ 2.5 & 10.3 $\pm$ 2.4 &
Delayed BRN & 33.0 $\pm$ 27.2 & 10.9 $\pm$ 9.4 & 8.7 $\pm$ 10.1 & 13.1 $\pm$ 9.4 & 9.1 $\pm$ 9.9 \\ \hline

SSBkg & 3889.0 $\pm$ 80.0 & 3825.8 $\pm$ 53.0 & 3810.5 $\pm$ 52.7 & 3826.8 $\pm$ 55.8 & 3807.8 $\pm$ 51.9 &
SSBkg & 3152.0 $\pm$ 24.1 & 3128.3 $\pm$ 23.6 & 3129.1 $\pm$ 24.7 & 3130.5 $\pm$ 22.6 & 3118.3 $\pm$ 24.3 \\ \hline

Total & 4288.0 $\pm$ 83.4 & 4199.5 $\pm$ 53.0 & 4204.5 $\pm$ 53.1 & 4180.6 $\pm$ 56.1 & 4179.5 $\pm$ 55.4 &
Total & 3807.9 $\pm$ 165.6 & 3834.9 $\pm$ 46.0 & 3865.9 $\pm$ 43.9 & 3867.3 $\pm$ 46.3 & 3841.2 $\pm$ 42.7 \\ \hline

\end{tabular}
}

%% file: tables/param_fit_data.tex
\caption{Resulting best-fit values of the parameters of interest $\mathcal{P}_i$, along with their uncertainties. The linear correlation coefficients $r_{ij}$ between parameters $\mathcal{P}_i$ and $\mathcal{P}_j$ are also provided. Rows correspond to specific analyses as marked in the first three columns.}
    \label{tab:param_fit}
\resizebox{\textheight}{!}{%
\begin{tabular}{|l|l|l|l|l|l|l|l|l|l|l|l|l|}
\hline
\textbf{Dataset}            & \textbf{Form factor} & \textbf{$\mathbf{R_n}$ Prior}        & \textbf{$\mathbf{\mathcal{P}_1}$} & \textbf{$\mathbf{\mathcal{P}_2}$} & \textbf{$\mathbf{\mathcal{P}_3}$} & \textbf{$\mathbf{\mathcal{P}_4}$}       & \textbf{$\mathbf{r_{12}}$} & \textbf{$\mathbf{r_{13}}$} & \textbf{$\mathbf{r_{14}}$} & \textbf{$\mathbf{r_{23}}$} & \textbf{$\mathbf{r_{24}}$} & \textbf{$\mathbf{r_{34}}$} \\ \hline
\multirow{3}{*}{CsI + l-Ar} & Helm                 & \multirow{9}{*}{+/- 5\%} & 0.857 ± 0.25  & -0.69 ± 0.23  & 730 ± 6.2e+02 & -2.79e+05 ± 3.5e+05 & -0.713            & -0.331            & 0.284             & -0.388            & 0.274             & -0.86             \\ \cline{2-2} \cline{4-13} 
                            & Sym. Fermi           &                          & 0.85 ± 0.25   & -0.688 ± 0.23 & 713 ± 6.1e+02 & -2.84e+05 ± 3.4e+05 & -0.72             & -0.35             & 0.296             & -0.361            & 0.252             & -0.858            \\ \cline{2-2} \cline{4-13} 
                            & Klein-Nystrand       &                          & 0.856 ± 0.25  & -0.69 ± 0.23  & 730 ± 6.2e+02 & -2.77e+05 ± 3.5e+05 & -0.716            & -0.337            & 0.289             & -0.38             & 0.275             & -0.868            \\ \cline{1-2} \cline{4-13} 
\multirow{3}{*}{CsI}        & Helm                 &                          & 1.03 ± 0.31   & -0.81 ± 0.38  & 495 ± 9.5e+02 & 3.74e+04 ± 5.7e+05  & -0.655            & -0.118            & 0.227             & -0.639            & -0.002            & -0.431            \\ \cline{2-2} \cline{4-13} 
                            & Sym. Fermi           &                          & 1.02 ± 0.3    & -0.789 ± 0.38 & 434 ± 9.4e+02 & 3.89e+04 ± 5.7e+05  & -0.665            & -0.108            & 0.229             & -0.637            & 0.001             & -0.444            \\ \cline{2-2} \cline{4-13} 
                            & Klein-Nystrand       &                          & 1.03 ± 0.31   & -0.807 ± 0.38 & 490 ± 9.5e+02 & 3.53e+04 ± 5.7e+05  & -0.652            & -0.126            & 0.225             & -0.636            & 0.0               & -0.433            \\ \cline{1-2} \cline{4-13} 
\multirow{3}{*}{l-Ar}       & Helm                 &                          & 0.767 ± 0.51  & -0.606 ± 0.85 & 632 ± 1.3e+03 & -3.03e+05 ± 4.2e+05 & -0.241            & -0.464            & 0.401             & -0.626            & 0.274             & -0.715            \\ \cline{2-2} \cline{4-13} 
                            & Sym. Fermi           &                          & 0.754 ± 0.51  & -0.619 ± 0.85 & 663 ± 1.3e+03 & -3.27e+05 ± 4.1e+05 & -0.241            & -0.456            & 0.401             & -0.624            & 0.275             & -0.708            \\ \cline{2-2} \cline{4-13} 
                            & Klein-Nystrand       &                          & 0.758 ± 0.51  & -0.61 ± 0.85  & 645 ± 1.3e+03 & -3.11e+05 ± 4.2e+05 & -0.237            & -0.457            & 0.391             & -0.623            & 0.29              & -0.709            \\ \hline
\multirow{3}{*}{CsI + l-Ar} & Helm                 & \multirow{9}{*}{Uniform} & 0.945 ± 0.28  & -0.719 ± 0.25 & 590 ± 7e+02   & -1.27e+05 ± 4.4e+05 & -0.694            & -0.312            & 0.285             & -0.39             & 0.225             & -0.84             \\ \cline{2-2} \cline{4-13} 
                            & Sym. Fermi           &                          & 0.932 ± 0.28  & -0.72 ± 0.25  & 627 ± 6.9e+02 & -1.72e+05 ± 4.2e+05 & -0.7              & -0.315            & 0.279             & -0.379            & 0.22              & -0.835            \\ \cline{2-2} \cline{4-13} 
                            & Klein-Nystrand       &                          & 0.945 ± 0.29  & -0.721 ± 0.25 & 600 ± 7.1e+02 & -1.39e+05 ± 4.3e+05 & -0.695            & -0.318            & 0.273             & -0.384            & 0.238             & -0.84             \\ \cline{1-2} \cline{4-13} 
\multirow{3}{*}{CsI}        & Helm                 &                          & 1.13 ± 0.34   & -0.937 ± 0.41 & 848 ± 1.1e+03 & 5.92e+04 ± 5.7e+05  & -0.681            & 0.014             & 0.191             & -0.69             & -0.054            & -0.282            \\ \cline{2-2} \cline{4-13} 
                            & Sym. Fermi           &                          & 1.12 ± 0.34   & -0.927 ± 0.41 & 812 ± 1.1e+03 & 5.48e+04 ± 5.7e+05  & -0.686            & 0.023             & 0.184             & -0.691            & -0.043            & -0.291            \\ \cline{2-2} \cline{4-13} 
                            & Klein-Nystrand       &                          & 1.13 ± 0.34   & -0.936 ± 0.41 & 844 ± 1.1e+03 & 5.84e+04 ± 5.7e+05  & -0.684            & 0.024             & 0.18              & -0.693            & -0.047            & -0.281            \\ \cline{1-2} \cline{4-13} 
\multirow{3}{*}{l-Ar}       & Helm                 &                          & 0.762 ± 0.51  & -0.586 ± 0.85 & 596 ± 1.3e+03 & -2.11e+05 ± 4.7e+05 & -0.232            & -0.455            & 0.353             & -0.62             & 0.232             & -0.66             \\ \cline{2-2} \cline{4-13} 
                            & Sym. Fermi           &                          & 0.763 ± 0.51  & -0.585 ± 0.85 & 592 ± 1.3e+03 & -2.26e+05 ± 4.6e+05 & -0.238            & -0.452            & 0.368             & -0.615            & 0.241             & -0.672            \\ \cline{2-2} \cline{4-13} 
                            & Klein-Nystrand       &                          & 0.756 ± 0.51  & -0.579 ± 0.84 & 597 ± 1.3e+03 & -2.19e+05 ± 4.7e+05 & -0.225            & -0.458            & 0.351             & -0.619            & 0.249             & -0.666            \\ \hline
\end{tabular}
}

%% file: sections/results.tex
The primary objective of the analysis is to obtain posteriors on the parameter set $\vec{\mathcal{P}}$ marginalized over the remaining uncertain parameters. Specific physical realizations of $\vec{\mathcal{P}}$ along the lines of Eqns.~\ref{eqn:p_decompose}~\&~\ref{eqn:xi_decompose} may then be used to transcribe these constraints to physical parameters, like the Weak mixing angle ($\sin^2\theta_W$), the Weak radius ($R_W$) or $\nu - q$ coupling strengths. These derived constraints would be subject to their respective physical realizations and nuclear response parameterizations. Posteriors on $\vec{\mathcal{P}}$ from a global analysis of $\cevns$ data from various target isotopes may be used to constrain, in particular, the common physical parameters entering the Weak response. Deviations in the posteriors across individual analyses of different targets may be used to understand residual isotope dependence, for instance, due to additional nuclear responses. Finally, these posteriors enable model-agnostic experimental extraction of the differential cross-section for a given nuclear isotope, which can then be reused across different flux and detector configurations. The posteriors naturally incorporate the impact of systematic uncertainties associated with the specific datasets used. Eqn.~\ref{eqn:rate-fit}, used with the posteriors and the other relevant parameters alongside their uncertainties, provides the observed differential cross-section based on the datasets analyzed. At the overwhelmingly dominant leading order, the posteriors from the global analysis are also expected to produce accurate predictions for $\cevns$ differential cross-sections for all targets.

In the first stage of the analysis, each dataset is analyzed individually. Each individual analysis is also repeated using the three different form-factor parameterizations mentioned in Sec.~\ref{sec:xsec}. Each of the above is also repeated using a uniform and normal prior on the Weak radius of the nuclei ($R_W$). In the second stage, all the datasets are used in a global analysis with similar variations on the choice of nuclear responses. The analysis was performed with the open-source Julia package {\it Newtrinos.jl}\cite{Eller:2026icp} previously used in Refs.~\cite{Ettengruber:2024fcq,Kozynets:2024xgt,Eller:2025lsh,Eller:2026urd}. The full analysis pipeline used in this work has also been added to this package as publicly usable modules. The full set of parameters along with their nominal values and prior distributions as used for each of the two datasets considered are listed in Table~\ref{tab:params}.

The full set of best-fit values of the normalization of each component, along with their uncertainties, for each analysis is summarized in Table~\ref{tab:norm_fit}. The prior numbers for the $\cevns$ norms are essentially their SM predictions, while those for the background components are sums over the event data provided by the Collaboration in its public data release. Results using the Helm parameterization of the Weak nuclear response along with a Normal prior on the Weak radius are referred to when reporting the global best-fit. The global best-fit distributions of the signal and background components for the COHERENT Ar and CsI datasets are shown in Fig.~\ref{fig:fit_spectrum} on each of the observable variables. In terms of the norms, the global best-fit values of the CE$\nu$NS signal component for the CsI dataset is $328.9 \pm 30.9$ within a window of $[5,200]$PE, which is consistent with the COHERENT collaboration's reported value of $306 \pm 20$ for a smaller analysis window~\cite{COHERENT:2021xmm}. The best-fit value of the CE$\nu$NS signal component for the l-Ar dataset is $156.0 \pm 43.3$ for a window of $[3,300]$keVnr, which is also consistent with the COHERENT collaboration's reported value of $159 \pm 43$~\cite{COHERENT:2020iec}. The best-fit values of the background components are also consistent with those reported by the COHERENT collaboration. Notably, purely in terms of the signal normalization, the CsI dataset observes fewer events than predicted by the SM while the l-Ar reports an excess. Although both are within 1$\sigma$ of the SM prediction, they show pulls in opposite directions across all analyses. This feature can also be observed in the results reported by the Collaboration.

The posteriors on the parameters of interest $\vec{\mathcal{P}}$ are shown in Fig.~\ref{fig:corner}. The lower panel of this figure shows their marginal distributions for each choice of dataset, nuclear response parameterization and prior distribution. The best-fit values for $\vec{\mathcal{P}}$, along with their uncertainties and linear correlation coefficients, are summarized in Table~\ref{tab:param_fit} for each choice. Posteriors on $\vec{\mathcal{P}}$ from the individual analyses are shown with colored contours, while the global ones are shown in gray. The l-Ar dataset corresponds to a much smaller net exposure than the CsI dataset. It also comes with much larger beam-coincident backgrounds. Therefore, it is expected {\it a priori} that the CsI dataset, being statistically favored, would lay much sharper constraints on the signal strength and shape. This is vindicated by the posteriors on $\vec{\mathcal{P}}$ in Fig.\ref{fig:corner}, but only partially. The posteriors on $\mathcal{P}_1$ and $\mathcal{P}_2$ are indeed much sharper for the CsI dataset, but the l-Ar dataset gains in relative sensitivity to $\mathcal{P}_3$ and it is more sensitive to $\mathcal{P}_4$ in comparison. As described in Sec.\ref{sec:xsec}, $\mathcal{P}_3$ and $\mathcal{P}_4$ are most strongly constrained by the high-$E_R$ spectrum. Larger nuclei, such as $^{133}$Cs and $^{127}$I, have a much larger loss of coherence at high $E_R$ than smaller nuclei such as Ar. Therefore, the impact of $\mathcal{P}_3$ and $\mathcal{P}_4$ is more strongly damped for CsI than for l-Ar. This explains why the l-Ar dataset has a stronger sensitivity to these sub-dominant parameters than the CsI dataset, despite being statistically disfavored. Even so, the l-Ar dataset has rather poor sensitivity overall, especially in the first two, dominant parameters. The combined analysis of both datasets provides the most stringent constraints on all parameters of interest, effectively exploiting the most sensitive datasets in each case.

The posteriors on $\vec{\mathcal{P}}$ for the two datasets do not exhibit a clear opposite pull, unlike their norms. This is especially true for $\mathcal{P}_1$ \& $\mathcal{P}_3$ where both datasets display similar behavior with respect to their SM expectations. The choice of $F_{\rm ref}$ also has nearly no impact on the best-fits and the distributions of the parameters. The choice of uniform or Normal $R_W$ priors, however, do affect the posterior distributions. Whether these observations indicate the expected coincidence of posteriors on $\vec{\mathcal{P}}$ across targets may be vindicated as more sensitive datasets become available. The posteriors on the different components are noticeably correlated to each other, although one must discount the fact that the l-Ar dataset has relatively weak resolving power. The CsI dataset displays marked negative correlation between the $\mathcal{P}_1$-$\mathcal{P}_2$ and $\mathcal{P}_2$-$\mathcal{P}_3$ pairs while being most sensitive to these parameters at the same time. The l-Ar dataset show similar strong correlation between the $\mathcal{P}_3$-$\mathcal{P}_4$ pair, which are also the parameters it is most sensitive to. The choice of $F_{\rm ref}$ does not affect the correlations. The choice of uniform or Normal priors on $R_W$ does affect the degree of correlation but does not change their nature or relative strength. The combined analysis shows similar strong correlation between the $\mathcal{P}_1$-$\mathcal{P}_2$ and $\mathcal{P}_3$-$\mathcal{P}_4$ pairs while the correlation between $\mathcal{P}_2$ \& $\mathcal{P}_3$, conspicuous in the CsI dataset, is much reduced.

The posteriors on $\vec{\mathcal{P}}$ alongside those on the relevant nuisance parameters may be directly utilized to constrain the $\cevns$ differential cross-section through Eqn.\ref{eqn:rate-fit}. The final best-fit differential cross-sections on $^{40}$Ar, $^{133}$Cs and $^{127}$I are shown in Fig.~\ref{fig:cevns_xsec} for a few sample neutrino energies. It shows the extent of variation in the $d\sigma/dE_R$ for each isotope that the two analyzed datasets may tolerate at a $68\%$ confidence level. The allowed extent is shown as shaded bands for both the individual and combined analyses. They are overlaid with the SM prediction in each case. The statistically favored CsI dataset has greater constraining power on the total differential cross-section than its l-Ar counterpart. These favored bands may be used directly as experimental extractions of the differential cross-section in future analyses.

%% file: sections/LZ_test.tex
The LUX-ZEPELIN (LZ) experiment has recently reported its results from a search for dark matter particles interacting with Xenon nuclei~\cite{LZ:2026axp}. Particularly, it reports having observed one event matching a nuclear recoil of $248\,\pm\,23\,(\text{stat})\,\pm\,23\,(\text{syst})\,\text{keV}$, inconsistent with the expected background. This test examines the possibility of it being an unlikely $\cevns$ event. The posteriors from the analyzed $\cevns$ datasets are used with Eqn.~\ref{eqn:rate-fit} to obtain the expected $\cevns$ cross-section in the LZ window, rather than the usual SM modeling.
Under this premise, the study evaluates the average available $\cevns$ differential cross-section over the reported recoil energy window for a range of possible neutrino energies.

The measured recoil energy was taken to be centered at
$E_r=248\,\text{keV}$, with statistical and systematic uncertainties of
$23\,\text{keV}$ each. These uncertainties were combined in quadrature,
and a fixed window of approximately $[215.5,280.5]\,\text{keV}$ was used.
For a Xenon isotope $i$, the differential CEvNS cross section is denoted by
$d\sigma_i/dE_r$. The natural-xenon differential cross section is the
isotopic-abundance-weighted sum:
\begin{equation}
  \frac{d\sigma_{\rm Xe}}{dE_r}(E_r,E_\nu)
  = \sum_i f_i \frac{d\sigma_i}{dE_r}(E_r,E_\nu),
\end{equation}
where $f_i$ is the natural abundance of isotope $i$. The effective average
differential cross-section in the LZ recoil window is defined as
\begin{equation}
  \left\langle\frac{d\sigma_{\rm Xe}}{dE_r}\right\rangle(E_\nu^\ast)
  = \frac{1}{E_r^+-E_r^-}
    \int_{E_r^-}^{E_r^+}
    \frac{d\sigma_{\rm Xe}}{dE_r}(E_r,E_\nu^\ast)\,dE_r
  \label{eqn:lz_effective_xsec}
\end{equation}
where $E_\nu^\ast$ is the incident neutrino energy. The expression holds for all $E_\nu^\ast$ that satisfy the kinematic condition of Eqn.~\ref{eqn:kinematic_cond}, and is zero otherwise. 
The full, fixed window is used for the averaging even when only part of that window is
kinematically accessible. Neutrino energies for which $\left\langle{d\sigma_{\rm Xe}}/{dE_R}\right\rangle>0$ are said to have positive support within the recoil window.

The differential cross-section is calculated using Eqn.~\ref{eqn:rate-fit} with the posterior draws of the $\cevns$ parameters. This requires the respective reference form factor and Weak charge to be plugged in separately, as used when obtaining the posteriors themselves. The Helm form factor was used in the present test, with the Weak radius $R_W$ for the respective Xenon isotopes being the only target-specific parameter modulating it. The Weak charge uses the Weak mixing angle ($\sin^2\theta_W$) as in the analysis that produced the posteriors. The proton and neutron numbers for the respective Xenon isotopes are used to produce the relevant Weak charges. To correctly account for all uncertainties in the cross-section, these parameters must be sampled from their respective prior or posterior distributions, as the case may be.
Since the target material in this case is Xenon, the posterior draws fit to the CsI dataset are considered to be a close approximation. The posteriors from the global fit are also expected to be sufficiently accurate for this test, as described in the previous sections. The test is performed for both cases, and the results are compared with the SM benchmark, with the parameters $\vec{\mathcal{P}}$ fixed at their SM values.

The draws were ordered by posterior density, and the highest-density batch containing $68\%$ of the
posterior weight was used for the support test. The full batch of posterior draws was used in each case in order to ensure that all correlations between the nuisance parameters and $\vec{\mathcal{P}}$ are preserved. The only source of uncertainty in the SM-benchmark test comes from the uncertainties in the Weak radii for each isotope and the Weak mixing angle.

\begin{figure}[t]
  \centering
  \includegraphics[width=\linewidth]{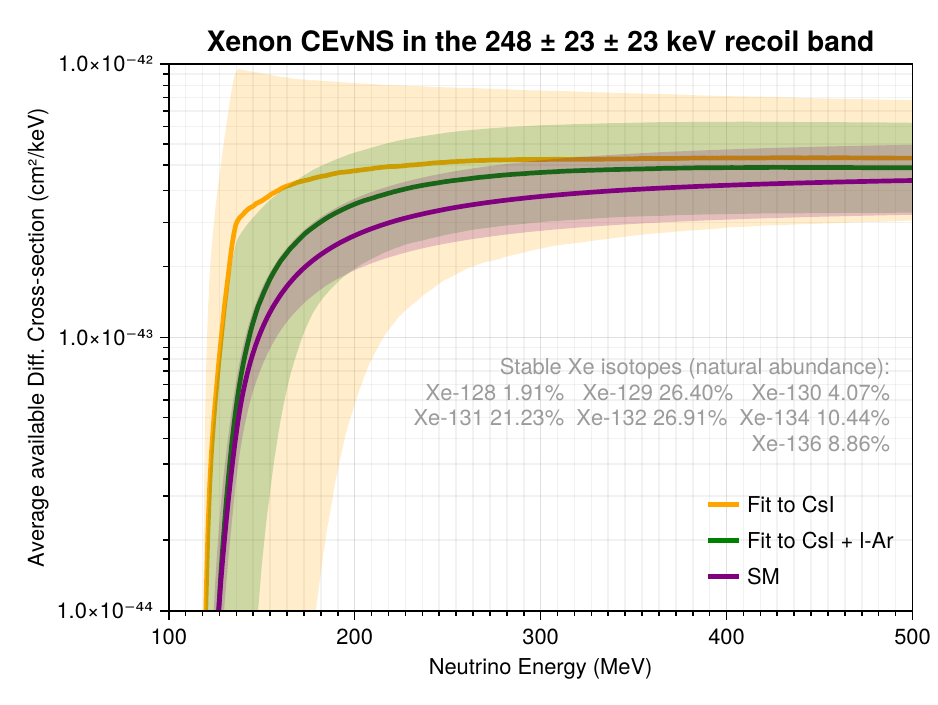}
  \caption{Effective average xenon CEvNS cross section in the LZ recoil
  window as a function of neutrino energy. The posterior mean is shown by the
  solid curve, and the central 68\% interval is shown
  by the shaded band.}
  \label{fig:lz-xe-effective-xsec}
\end{figure}

The result is expressed as a function of incident neutrino energy in Fig.~\ref{fig:lz-xe-effective-xsec}, which shows the posterior mean and the central 68\% interval for each of the test cases, alongside the SM benchmark. It denotes the average available differential cross-section for a $\cevns$ process to produce a nuclear recoil within the reported LZ recoil window as a function of incident neutrino energy, based on observed $\cevns$ spectra. The minimum neutrino energy with positive support coincides with the kinematic threshold of $\sim 120$ MeV. High-$E_\nu$ suppression of $\cevns$ is driven by the high-energy cutoff in the neutrino flux, when present. This test is conducted at the level of the differential cross-section; hence, such suppression is not explicit. However, Eqn.~\ref{eqn:lz_effective_xsec} could also be interpreted as the flux-integrated differential rate for a mono-energetic neutrino flux, sans the normalizing factors for target and neutrino flux densities. Coherence (or the lack of it) depends on the transferred momentum and nuclear characteristics, as discussed in Sec.~\ref{sec:cross_section}. A relatively high-energy neutrino could, therefore, still produce a coherent response within the support window for the reported recoil energies.

The available cross-section remains quite low due to loss of coherence over the considered recoil energy window, typically suppressed by a factor of $\lesssim 10^{-3}$ relative to full coherence. The bands of support from the CsI and the global posteriors are generally broader than their SM counterparts. This is as expected since they incorporate additional uncertainties from the data. The CsI dataset has the broadest posterior distributions among the test cases, and the corresponding limits of support are the broadest. Both the CsI-only and global-fit posteriors also allow for a higher available cross-section compared to the SM, as visible in Fig.~\ref{fig:lz-xe-effective-xsec} from the solid lines tracing the result from posterior means. These mean values gradually merge near the kinematic threshold at low neutrino energies, and likewise asymptotically at higher energies. The 68\% intervals also seem to gradually coincide as one tests for higher neutrino energies.

These results may be translated into a more realistic expected integrated event rate by applying appropriate normalizations to the target exposure and the neutrino flux. This is in addition to accounting for all detector effects mapping the physical nuclear recoil energies to the observed detector response, including the corresponding uncertainties. Alternative flux hypotheses may also be used in Eqn.~\ref{eqn:lz_effective_xsec} for a more exhaustive examination. The hypotheses of a single neutrino producing the reported event may not be treated at par with that of a continuous neutrino flux. The latter would, in the best case, produce a single event despite the predicted differential cross-section being a shape over the recoil energies. The predicted rate of events in that case may also depend on the directionality of the neutrino and the detector geometry. A multi-bin Poisson fit may also not be the most appropriate method for testing the hypothesis in that case. This comprehensive treatment of the reported results deserves a separate dedicated study, which is beyond the scope of this work. 

%% file: sections/summary.tex
Having vindicated the Standard Model (SM) prediction of Coherent Elastic Neutrino-Nucleus Scattering ($\cevns$) with its first observation in 2017 by the COHERENT Collaboration, the frontier now lies in its precise characterization. $\cevns$ is unique in that its response is sensitive to three separate scales of physical inputs entangled non-trivially and dependent sequentially, starting from the shortest to the longest length scales. The first layer at the shortest length scales are the neutrino-quark coupling strengths, which are sensitive to the Weak mixing angle and potential beyond-the-Standard-Model (BSM) physics. The second layer incorporates physical inputs at the nucleonic level characterized by hadronic matrix elements, the specific nature of which depends on the form of $\nu-q$ current under consideration. The third layer, at the longest length scale, is the nuclear response, specific to the nature of the former layers and dependent on the target nucleus in question. Deviations from expected behavior in any of these inputs manifest as variations in both the shape and normalization of the expected $\cevns$ spectrum.
While this rich interplay provides a unique opportunity to probe physics at multiple scales through a single process, it also makes a detailed interpretation of $\cevns$ measurements highly challenging. At a predictive level, these three layers cannot be factorized except under several simplifying assumptions or within constricted kinematic regimes. Disentangling these layers requires a global analysis of multiple $\cevns$ observations from complementary targets, neutrino source profiles and detector sensitivities.
This work presents a general framework for the analysis of the $\cevns$ spectrum, enabling a numerical extraction of its differential cross-section from experimental data and a global interpretation of the resulting constraints in terms of the underlying nuclear and particle physics. The framework has been implemented in the open-source public \texttt{Newtrinos.jl} package.

The $\cevns$ differential cross-section is parametrized on a basis, $\vec{\mathcal{B}}$, polynomial in the neutrino energy ($E_\nu$), nuclear recoil energy ($E_R$) and target nuclear mass ($M_N$). The choice is motivated by the kinematic forms common to a broad range of SM and BSM scenarios. Multipliers, $\vec{\mathcal{P}}$, are assigned to the basis elements to provide a flexible parameterization of the cross-section's shape and normalization. In the SM, these multipliers take constant numerical values at leading order, with subleading corrections that typically scale as $\sim N^{-2}$, where $N$ is the neutron number of the target. The dominant Weak response for each target species is factorized into a reference Weak nuclear form factor, $F_{\mathrm{ref}}(q^2;R_W)$, and the SM Weak charge, $Q_{\mathrm{W,SM}}$, with the Weak radius, $R_W$, and the Weak charge treated as nuisance parameters. The $\cevns$ response for each isotope is evaluated independently and summed, weighted by their respective abundances in the target.

The forward model separates the physical differential cross-section, neutrino flux, and detector response into modular components, each with its own parameters and associated prior distributions. This construction permits alternative implementations of individual components and allows any model parameter to be studied as a parameter of interest. In the analyses presented here, $\vec{\mathcal{P}}$ is inferred through Bayesian fits with uninformed priors. The resulting posterior distributions provide a numerical, model-agnostic extraction of the differential cross-section, with theoretical and experimental uncertainties propagated through the forward model. They can subsequently be combined with alternative neutrino flux and detector response models to predict event rates in other experiments. Conversely, all physical models whose predictions can be expressed on the basis $\vec{\mathcal{B}}$ can be tested against the inferred posterior distributions.

The inferred multipliers also enable the interpretation of the extracted cross-section in terms of the underlying nuclear and particle physics. Within the SM, $\vec{\mathcal{P}}$ can be expressed in terms of the Weak and Axial nuclear form factors, which depend on nuclear response functions, hadronic matrix elements, and the Weak mixing angle. Multiple explicit evaluations of these form factors using all of these inputs are available in the literature, utilizing different experimental inputs and theoretical assumptions. Once a choice of these form factor implementations is made, the interpretation of all the components of $\vec{\mathcal{P}}$ in their terms can be compared with the full posterior distribution over these parameters to constrain the Weak mixing angle. In BSM interpretations, the same approach can be used to constrain the Weak charge of a target nucleus. The Weak radius can likewise be constrained through its relation to the Weak charge and the relevant hadronic matrix elements. These interpretations are target-specific and implementation-dependent, and the resulting constraints apply to the momentum-transfer range probed by the datasets.

The isotope-resolved formulation accommodates targets with multiple nuclear species by evaluating each isotope's response separately and combining the contributions weighted by their abundances. The inferred posterior distributions constrain deviations of the relevant nuclear responses from the reference weak response that remain compatible with the observed recoil spectrum. This interpretation is most direct for single-isotope targets or targets composed of closely related nuclear species. Where sufficiently informative measurements are available, the resulting constraints may reduce the need for additional nuclear-response uncertainties in subsequent analyses. This is relevant to dark-matter direct-detection experiments, where $\cevns$ contributes to the background, and to neutrino-oscillation experiments, where uncertainties in nuclear responses can be an important source of systematic error.

The framework is applied to two datasets, COHERENT CsI and liquid argon (l-Ar), with the detailed results reported in this article. The CsI dataset generally provides greater statistical sensitivity, while the two datasets exhibit complementary sensitivities to different components of $\vec{\mathcal{P}}$. Their combination is therefore shown to provide constraints inaccessible from either dataset alone, illustrating the importance of complementary targets in global analyses.
As an illustrative application, the framework is used to examine the recently reported event at the LUX-ZEPELLIN (LZ) experiment. The experimentally extracted $\cevns$ differential cross-section is used to assess the cross-section available within the corresponding nuclear-recoil window. The resulting range of available cross-section is small but nonzero and slightly larger than predicted by the SM. This illustrates the framework's applicability for evaluating the potential $\cevns$ contribution within a specific experimental recoil range.

More generally, the explicit separation of interaction physics, neutrino flux, and detector response enables assessment of the sensitivity of different experimental configurations in terms of both the recoil spectrum and the underlying physical parameters. This provides a basis for selecting complementary targets and neutrino sources, identifying detector improvements that enhance sensitivity to particular components of the cross-section, and combining datasets in global analyses. Such analyses can improve parameter constraints and help disentangle nuclear, hadronic, and particle-physics contributions to $\cevns$. The framework therefore provides a unified approach to extracting the differential cross-section from experimental data, interpreting the results within specific physical models, and informing the design of future measurements.

%% file: sections/COHERENT_Ar.tex
\input{tables/lar_response_params}
\input{tables/lar_response_components}
For the liquid-argon dataset, we follow the published COHERENT analysis pathway and dataset~\cite{COHERENT:2020iec,Akimov:2020czh}. Nominal values of parameters and their priors are taken from the released parameter file \texttt{LArParametersAnlA.yaml}. The target is treated as a single effective $^{40}$Ar nucleus, so that the isotope dependence enters only through one recoil spectrum and one neutron-radius parameter in the cross-section sector. The detector response is then built on top of this recoil prediction. Internally, the recoil-energy axis is discretized over $3$--$300\,$keV$_{\rm nr}$ in $0.5\,$keV$_{\rm nr}$ bins, after which the nuclear-recoil spectrum is mapped to the electron-equivalent observable through an explicit quenching prescription. 

The default neutrino flux is the Analytic branch (with distance $L_{\rm Ar}=2750\,$cm and exposure $X_{\rm Ar}=6.12\,$GWh). The corresponding flux normalization uncertainty is carried by
\begin{equation}
    r=0.09\pm0.009,
\end{equation}
implemented as a truncated normal prior on $[0.05,0.15]$. The analytic branch does not have a timing structure built-in and is therefore not sensitive to the timing nuisance parameter $\delta t_{\rm flux}$. As a result, the timing information for the l-Ar dataset must be incorporated through the timing data provided with Ref.~\cite{COHERENT:2020iec} at the detector response level.

The quenching factor is written as a low-order polynomial,
\begin{equation}
	E_{\rm ee} = \left(a+bE_R\right)E_R,
\end{equation}
with the nuisance parameters factorized as
\begin{equation}
	a = a_0 + \delta a,\qquad b = b_0 + \delta b,
\end{equation}
where $(a_0,b_0)=(0.246,7.8\times 10^{-4}\,{\rm keV}_{\rm nr}^{-1})$ and the shifts are constrained by the quoted Gaussian uncertainties. This construction preserves a transparent link between the physical recoil spectrum and the experimentally reconstructed energy scale.

The detector resolution is modeled by a Gaussian kernel in the reconstructed-energy variable,
\begin{equation}
	R_E(i|E_R)\propto \epsilon(E_{\rm ee})\,\exp\left[-\frac{(E^{\rm out}_i-E_{\rm ee})^2}{2\sigma^2(E_{\rm ee})}\right],
\end{equation}
with width
\begin{equation}
	\sigma(E_{\rm ee}) = 0.58\sqrt{E_{\rm ee}},
\end{equation}
while the overall acceptance $\epsilon(E_{\rm ee})$ is obtained by linear interpolation of the published file \texttt{CENNS10AnlAEfficiency.txt}. The resulting response matrix therefore performs the simultaneous smearing and efficiency weighting needed to convert a binned recoil spectrum into predicted counts in the reconstructed-energy observable.
The effective active mass is floated in the fit,
\begin{equation}
	M_{\rm fid} = M_{\rm fid}^{(0)} + \delta M_{\rm fid},
\end{equation}
so that the overall signal normalization carries the quoted detector-mass uncertainty directly rather than through an external rescaling.

The liquid-argon data product contains more structure than reconstructed energy alone. In the default configuration used here, where the signal is generated from the analytic SNS flux branch, the predicted $\cevns$ spectrum after energy smearing is distributed over the published $F_{90}$ and timing observables through their corresponding normalized experimental templates. This effectively factorizes the detector description into a reconstructed-energy response and two auxiliary probability densities, one for pulse-shape discrimination and one for timing. If instead the time-resolved flux tables are used, the timing axis is inherited directly from the flux model and only the $F_{90}$ distribution is applied as an additional factor. This distinction allows the same detector prescription to accommodate either a one-dimensional or a two-dimensional signal input without changing the remainder of the likelihood.

Symbolically, the two branches may be written as
\begin{eqnarray}
	N^{\rm obs}_{i\ell m} &\propto& p_{F90}(\ell)\,p_t(m)\sum_j R_E(i|E_{R,j})N^{\rm phys}_j,\nn
	N^{\rm obs}_{i\ell m} &\propto& p_{F90}(\ell)\sum_{jk} R_E(i|E_{R,j})N^{\rm phys}_{jk},
\end{eqnarray}
for the analytic-flux and time-resolved-flux configurations respectively. Here $p_{F90}$ is extracted from \texttt{f90data1d.txt} and $p_t$ from \texttt{timingdata1d.txt}.

The uncertain detector-side parameters entering the l-Ar likelihood are therefore the quenching coefficients, the effective detector mass, and the normalizations of the prompt beam-related neutron, delayed beam-related neutron, and steady-state background components. In the present implementation, these are assigned Gaussian or truncated-Gaussian priors centered on the experimental nominal values, with relative widths matching the quoted uncertainties used in the analysis configuration. The detector-response ingredients are summarized in Table~\ref{tab:lar_response_components}, while the floated detector-side parameters used in the likelihood are listed in Table~\ref{tab:lar_response_params}.

%% file: tables/lar_response_params.tex
\begingroup
\squeezetable
\begin{table*}[t]
\caption{COHERENT l-Ar detector-side nuisance parameters, nominal values, and priors used in the fit model.}
\label{tab:lar_response_params}
{\small
\begin{tabular}{p{0.22\linewidth}p{0.18\linewidth}p{0.20\linewidth}p{0.30\linewidth}}
\hline
Parameter & Symbol & Nominal & Prior \\
\hline
QF slope & $a$ & $0.246$ & $\mathcal{N}(0.246,0.006)$ \\

QF curvature & $b$ & $7.8\times10^{-4}$ & $\mathcal{N}(7.8\times10^{-4},9\times10^{-5})$ \\

Fiducial mass & $M_{\rm fid}$ & $24.4\,$kg & truncated normal, $\sigma=0.61\,$kg \\

Prompt BRN norm & $N_{\rm pBRN}$ & $\sum$ \texttt{brnpdf} & truncated normal, $32\%$ \\

Delayed BRN norm & $N_{\rm dBRN}$ & $\sum$ \texttt{delbrnpdf} & truncated normal, $100\%$ \\

SS norm & $N_{\rm SS}$ & $\sum$ \texttt{bkgpdf} & truncated normal, $0.8\%$ \\
\hline
\end{tabular}
}
\end{table*}
\endgroup

%% file: tables/lar_response_components.tex
\begingroup
\squeezetable
\begin{table*}[t]
\caption{COHERENT l-Ar detector-response components used in the implementation.}
\label{tab:lar_response_components}
{\small
\begin{tabular}{p{0.27\linewidth}p{0.45\linewidth}p{0.22\linewidth}}
\hline
Component & Symbolic form & Source file \\
\hline
Efficiency & $R_E(i|E_R)$ & \texttt{CENNS10AnlAEfficiency.txt} \\

$F_{90}$ PDF & $p_{F90}(\ell)$ & \texttt{f90data1d.txt} \\

Timing PDF & $p_t(m)$ & \texttt{timingdata1d.txt} \\

Prompt BRN template & $B_{\rm pBRN}$ & \texttt{brnpdf.txt} \\

Delayed BRN template & $B_{\rm dBRN}$ & \texttt{delbrnpdf.txt} \\

SS background & $B_{\rm SS}$ & \texttt{bkgpdf.txt} \\
\hline
\end{tabular}
}
\end{table*}
\endgroup

%% file: sections/COHERENT_CsI.tex
\input{tables/csi_response_components}
\input{tables/csi_response_params}
The CsI dataset is intrinsically two-dimensional, combining pulse-height and timing information. Accordingly, its detector-response prescription is constructed as the product of an energy-to-photoelectron response and an explicit timing acceptance, following the structure specified in the COHERENT CsI release. The target is treated as a two-isotope mixture of Cs and I with their physical isotopic fractions, and the flux-folded recoil spectra from the two isotopes are combined before detector effects are applied. The recoil grid spans $3$--$200\,$keV$_{\rm nr}$ in $0.5\,$keV$_{\rm nr}$ steps, while the observed pulse-height axis is represented in the non-uniform PE binning used in the experimental analysis.

The default neutrino flux implementation uses the Data branch (with distance $L_{\rm CsI}=1960\,$cm and exposure $X_{\rm CsI}=13.99\,$GWh in the dataset assets). Two flux nuisance parameters are propagated in the fit: a global flux normalization,
\begin{equation}
    f_{\rm norm}\sim \mathcal{N}(1,0.1)\ \text{truncated to }[0.7,1.3],
\end{equation}
and a common timing offset,
\begin{equation}
    \delta t_{\rm flux}\sim \mathcal{N}(0,100\,{\rm ns})\ \text{truncated to }[-500,500]\,{\rm ns}.
\end{equation}
The imported tables are read as two-dimensional arrays over energy and time, converted to bin-integrated weights, rebinned to the analysis timing structure, and stored on MeV-based energy grids for the later rate integration.

The mapping from nuclear recoil to observable pulse height proceeds through a quenching prescription followed by stochastic smearing. In the implementation, the quenching relation is represented by a quartic polynomial in recoil energy,
\begin{equation}
	E_{\rm ee}(E_R;z_0,z_1) = 10^3\sum_{n=1}^4 q_n(z_0,z_1)x^n,\qquad x\equiv E_R/{\rm MeV}_{\rm nr},
\end{equation}
with coefficient vector factorized as
\begin{equation}
	\vec q(z_0,z_1) = \vec q^{\,(0)} + z_0\,\Delta\vec q^{\,(0)} + z_1\,\Delta\vec q^{\,(1)}.
\end{equation}
The nominal coefficients and deformation modes are extracted from \texttt{scintRespCoefficients.txt}, and the dimensionless nuisance amplitudes $z_0$ and $z_1$ are assigned truncated standard-normal priors. This provides a compact way of propagating the dominant uncertainty in the recoil-to-light conversion without overparameterizing the response. Once the mean electron-equivalent energy has been obtained, it is converted to a PE-distribution through a gamma-distributed response kernel,
\begin{equation}
	R_{\rm PE}(i|E_R) \propto \epsilon_{\rm PE}(\text{PE}_i;z_\epsilon)\int_{\Delta_i} d\text{PE}\,\Gamma(\text{PE};k,\theta),
\end{equation}
with
\begin{equation}
	k=1+bE_{\rm ee},\qquad \theta=\left[\frac{a}{E_{\rm ee}}\left(1+bE_{\rm ee}\right)\right]^{-1},
\end{equation}
where $(a,b)=(0.0749,9.56)$ are the resolution parameters used in the code. In this way, each recoil-energy bin contributes a normalized probability distribution over PE rather than a single reconstructed value.

Selection efficiency is then applied directly in PE space through a smooth sigmoid-like response function,
\begin{equation}
	\epsilon_{\rm PE}(\text{PE};z_\epsilon) = \frac{a_\epsilon(z_\epsilon)}{1+\exp\left[-b_\epsilon(z_\epsilon)(\text{PE}-c_\epsilon(z_\epsilon))\right]} + d_\epsilon(z_\epsilon),
\end{equation}
with coefficient vector
\begin{equation}
	\vec e(z_\epsilon)=\vec e^{\,(0)} + z_\epsilon\,\Delta\vec e,
\end{equation}
and $\vec e^{\,(0)}$, $\Delta\vec e$ extracted from \texttt{effCoefficients.txt}. The timing response is treated separately. The signal timing bins are inherited from the SNS flux module, and an explicit timing-efficiency function is applied on a per-bin basis. In continuous form, the prescription is
\begin{equation}
	\epsilon_T(t) = \left\{\begin{array}{ll}
		1, & t\le a_T,\\
		\exp\left[-b_T(t-a_T)\right], & t>a_T,
	\end{array}\right.
\end{equation}
with $(a_T,b_T)=(520\,{\rm ns},0.0494\times 10^{-3}\,{\rm ns}^{-1})$, after which the bin-averaged timing efficiencies are assembled into a diagonal matrix $R_T$. The full CsI detector response therefore factorizes into a PE response matrix acting on the recoil axis and a timing-response matrix acting on the beam-time axis,
\begin{equation}
	N^{\rm obs}_{i\alpha} = \sum_{j\beta} R_{\rm PE}(i|E_{R,j})\,R_T(\alpha|t_\beta)\,N^{\rm phys}_{j\beta}.
\end{equation}

The background model is built on the same observable space. The steady-state background is histogrammed directly from the beam-off data, while the prompt and delayed beam-related neutron components are constructed from the published PE and timing projections, rebinned to the common analysis binning, and combined into two-dimensional templates,
\begin{equation}
	B_X(i,\alpha) = N_X\,p_X^{\rm PE}(i)\,p_X^t(\alpha),\qquad X\in\{{\rm BRN,NIN}\},
\end{equation}
followed by the timing acceptance. The global timing-shift parameter carried by the SNS flux model is propagated to these beam-related backgrounds as well, so that any inferred offset in the beam structure is treated coherently for signal and background together. In addition, independent normalization nuisances are assigned to the steady-state, BRN, and NIN contributions.

The uncertain detector-response parameters entering the CsI likelihood are thus the quenching-shape deformations, the efficiency deformation, and the background normalizations, together with the shared timing-shift parameter inherited from the flux sector. This is precisely what makes the CsI detector model especially constraining: both spectral and temporal distortions are retained and propagated through a fully binned detector-response prescription before comparison with the observed data.
The detector-response ingredients are summarized in Table~\ref{tab:csi_response_components}, while the floated detector-side parameters used in the likelihood are listed in Table~\ref{tab:csi_response_params}.

%% file: tables/csi_response_components.tex
\begingroup
\squeezetable
\begin{table*}[t]
\caption{COHERENT CsI-2020 detector-response components used in the implementation.}
\label{tab:csi_response_components}
{\small
\begin{tabular}{p{0.27\linewidth}p{0.45\linewidth}p{0.22\linewidth}}
\hline
Component & Symbolic form & Source file \\
\hline
Quenching model & $E_{\rm ee}(E_R;z_0,z_1)$ & \texttt{scintRespCoefficients.txt} \\

PE efficiency & $\epsilon_{\rm PE}(\text{PE};z_\epsilon)$ & \texttt{effCoefficients.txt} \\

PE-time data & $N^{\rm obs}_{i\alpha}$ & \texttt{dataBeamOnC.txt} \\

SS background & $B_{\rm SS}$ & \texttt{dataBeamOnAC.txt} \\

BRN template & $p^{\rm PE}_{\rm BRN}\otimes p^t_{\rm BRN}$ & \texttt{brnPE.txt}, \texttt{brnTrec.txt} \\

NIN template & $p^{\rm PE}_{\rm NIN}\otimes p^t_{\rm NIN}$ & \texttt{ninPE.txt}, \texttt{ninTrec.txt} \\
\hline
\end{tabular}
}
\end{table*}
\endgroup

%% file: tables/csi_response_params.tex
\begingroup
\squeezetable
\begin{table*}[t]
\caption{COHERENT CsI-2020 detector-side nuisance parameters, nominal values, and priors used in the fit model.}
\label{tab:csi_response_params}
{\small
\begin{tabular}{p{0.20\linewidth}p{0.16\linewidth}p{0.21\linewidth}p{0.33\linewidth}}
\hline
Parameter & Symbol & Nominal & Prior \\
\hline
Efficiency mode & $z_\epsilon$ & $0$ & truncated $\mathcal{N}(0,1)$ \\

QF mode 0 & $z_0$ & $0$ & truncated $\mathcal{N}(0,1)$ \\

QF mode 1 & $z_1$ & $0$ & truncated $\mathcal{N}(0,1)$ \\

BRN norm & $N_{\rm BRN}$ & $\sum B_{\rm BRN}$ & truncated normal, $25\%$ \\

NIN norm & $N_{\rm NIN}$ & $\sum B_{\rm NIN}$ & truncated normal, $36\%$ \\

SS norm & $N_{\rm SS}$ & $\sum B_{\rm SS}$ & truncated normal, $2.1\%$ \\

Timing shift & $\delta t$ & $0$ & from flux sector \\
\hline
\end{tabular}
}
\end{table*}
\endgroup